\documentclass[pra,aps,floatfix,floats,superscriptaddress,notitlepage]{revtex4-1}

\pdfoutput=1

\usepackage{epsfig}
\usepackage{enumerate}
\usepackage{amsmath}
\usepackage{amssymb}
\usepackage{hyperref}
\usepackage{color}
\usepackage{theorem}
\usepackage{braket}
\usepackage{mathtools}
\usepackage{comment}
\usepackage{tikz}
\usepackage{empheq}
\usepackage{hyperref}

\newcommand{\be}{\begin{equation}}
\newcommand{\ee}{\end{equation}}
\newcommand{\bea}{\begin{eqnarray}}
\newcommand{\eea}{\end{eqnarray}}

\newcommand{\dd}{{\rm d}}
\definecolor{violet}{rgb}{0.62,0,1}
\definecolor{lightblue}{rgb}{0.12,0.56,1}
\definecolor{green}{rgb}{0.13,0.55,0.13}

\definecolor{violet}{rgb}{0.62,0,1}
\definecolor{lightblue}{rgb}{0.62,0,1}
\definecolor{green}{rgb}{0.13,0.55,0.13}

\def\doi{http://dx.doi.org/}

\def\eqref#1{(\ref{#1})}

\def\doi{http://dx.doi.org/}

\begin{document}

\title{From the sinh-Gordon field theory to the one-dimensional Bose gas:\\
	 exact local correlations and full counting statistics}

\author{Alvise Bastianello}
\author{Lorenzo Piroli}
\address{SISSA and INFN, via Bonomea 265, 34136, Trieste, Italy}

\begin{abstract}
We derive exact formulas for the expectation value of local observables in a one-dimensional gas of bosons with point-wise repulsive interactions (Lieb-Liniger model). Starting from a recently conjectured expression for the expectation value of vertex operators in the sinh-Gordon field theory, we derive explicit analytic expressions for the one-point $K$-body correlation functions $\langle (\Psi^\dagger)^K(\Psi)^K\rangle$ in the Lieb-Liniger gas, for arbitrary integer $K$. These are valid for all excited states in the thermodynamic limit, including thermal states, generalized Gibbs ensembles and non-equilibrium steady states arising in transport settings. Our formulas display several physically interesting applications: most prominently, they allow us to compute the full counting statistics for the particle-number fluctuations in a short
interval. Furthermore, combining our findings with the recently introduced generalized hydrodynamics, we are able to study multi-point correlation functions at the Eulerian scale in non-homogeneous settings. Our results complement previous studies in the literature and provide a full solution to the problem of computing one-point functions in the Lieb-Liniger model.
\end{abstract}

\maketitle

\section{Introduction}
\label{sec:intro}

Correlation functions encode all of the information which can be experimentally extracted from a many-body quantum system. At the same time, the problem of their computation is extremely complicated from the theoretical point of view, restricting us, in general, to rely uniquely on perturbative or purely numerical methods. 

An outstanding exception to this picture are integrable systems \cite{baxter-82}, characterized by the existence of an extensive number of local conservation laws, which provide an ideal theoretical laboratory to deepen our knowledge of many-body physics. This is especially true due to the possibility of obtaining \emph{exact}, unambiguous predictions for several quantities of interest, allowing us, for instance, to test the validity of approximate or numerical methods which can be applied to more general cases. While integrability directly provides the tools for diagonalizing the Hamiltonian, the computation of correlation functions constitute a remarkable challenge, which has attracted a constant theoretical effort over the past fifty years \cite{korepin,JimboBOOK,Thac81,efgk-05}. Classical studies have in particular focused on ground-state and thermal correlations, and joint efforts have led to spectacular results, for example in the case of prototypical interacting spin models such as the well-known Heisenberg chain \cite{JMMN92,MaSa96,KiMT00,GoKS04,CaMa05,BJMS07,TrGK09}.

More recently, new energy has been pumped into the study of integrable models, also due to the new experimental possibilities offered by cold-atom physics. Nearly ideal integrable systems can now be realized in cold-atom experiments both in and out equilibrium \cite{BlDZ08,PSSV11,CCGO11}, elevating the relevance of existing works beyond the purely theoretical interest, and motivating further advances in the framework of non-equilibrium physics (see \cite{CaEM16} for a collection of recent reviews on this topic). 

From the experimental point of view, one of the most relevant systems is the so-called Lieb-Liniger (LL) model \cite{LiLi63}. It describes a one-dimensional gas of point-wise interacting bosons, which can be realized in cold-atom experiments \cite{KiWW04,KiWW05,KiWW06,AEWK08,FPCF15}. While several results have already been obtained in the ground-state \cite{JiMi81,OlDu03,GaSh03,CaCa06,ChSZ05,ChSZ06,ScFl07,PiCa16-1} and at thermal equilibrium \cite{MiVT02,KGDS03,SGDV08,KuLa13,PaKl13,ViMi13,PaCa14,NRTG16}, the problem of computing its experimentally measurable correlation functions \cite{KiWW05,AJKB10,JABK11,hrm-11,FPCF15} for generic macrostates of the system still challenges the community. Until recently, even the simplest one-point functions appeared to be an open issue in the case of generic excited states. Even more urgent is the question on the \emph{full counting statistics} of local observables, most prominently for the particle-number fluctuations. Indeed, the latter provides fundamental information on the quantum fluctuations of the system, and can also be probed experimentally \cite{AJKB10,JABK11,HLSI08,KPIS10,KISD11,GKLK12}. Yet, no theoretical prediction for this quantity, not even approximate, was available in the existing literature for the Lieb-Liniger model. More generally, the full counting statistics of local observables in and out of equilibrium have been considered in many studies \cite{lp-08,gadp-06,cd-07,sk-13,lla-15,lddz-15,hb-17, sfr-11,cmv-12,ia-13,MoSC16, GuMW18,SuIv12,NiSE18}, even though analytical results in integrable systems have been provided only in a handful of cases~\cite{Eisl13,EiRa13, k-14,nr-17,CoEG17,sp-17,GrEC18}.

Recently, important progress on the problem of computing one-point functions in the one-dimensional Bose gas has been made, boosted by the results of Ref. \cite{KoMT09}, where a novel field-theoretical approach was introduced: the latter is based on the observation that the Lieb-Liniger model can be obtained as an appropriate non-relativistic (NR) limit of the sinh-Gordon (shG) field theory. In turn, one-point functions in this relativistic field theory can be obtained by means of the well-known LeClair-Mussardo series \cite{LeMu99}, which was exploited in Ref. \cite{KoMT09} to derive explicit formulas in the Lieb-Liniger gas. The ideas introduced in \cite{KoMT09} led to exact expressions for the experimentally relevant pair and three-body correlations \cite{KoCI11}, and were later fruitfully applied in the study of other models and field theories \cite{KoMP10,KoMT11,CaKL14,BaLM16,BaLM17}. Importantly, these results hold for arbitrary excited states, since the LeClair-Mussardo series itself was proven to be valid in general and not only for ground and thermal states \cite{Pozs11_mv}.

The findings of \cite{KoMT09} were later recovered and generalized by Pozsgay in Ref. \cite{Pozs11}. By exploiting a scaling limit of the XXZ Heisenberg chain to the Lieb-Linger gas \cite{GoHo87}, exact multiple-integral formulas were obtained for the generic $K$-body one-point function $\langle (\Psi^\dagger)^K(\Psi)^K\rangle$. Despite their conceptual importance, multiple-integral representations are not suitable for numerical evaluation. While this result could not be simplified further for generic $K$, it was possible to reach a simple integral expressions for $K=2,3,4$. We stress again that these formulas have been already applied to compute correlations in generic macrostates, including generalized Gibbs ensembles (GGEs) \cite{RDYO07,ViRi16,EsFa16,DWBC14}, which capture the long-time limit of the local properties of the system after a quantum quench \cite{cc-06}.

In this work we make a step forward and provide general formulas for the $K$-body one-point functions in the Lieb-Liniger model which are sufficiently simple to be easily evaluated numerically. Their form differs from the one found in \cite{Pozs11}, and only involves simple integrals. Our strategy follows the method introduced in \cite{KoMT09}: however, while the starting point of \cite{KoMT09} was provided by the LeClair-Mussardo series, we consider an alternative formula which has been recently conjectured by Negro and Smirnov \cite{NeSm13,Negr14} and later simplified in \cite{BePC16}. The latter provides an explicit resummation of the LeClair-Mussardo series in the case of a particular class of observables called vertex operators. Most of our results were previously announced in \cite{BaPC18}; here we present a detailed derivation, reporting in particular all the necessary technical calculations, the analytical and numerical checks, and a thorough discussion of the physical applications. In particular, in addition to the analysis of correlations in thermal and GGE states, we also discuss the implications of our findings for the full counting statistics of the number of particles in a small interval. Finally, within the framework of the recently introduced Generalized Hydrodynamics (GHD) \cite{CaDY16,BCDF16}, we present results for correlations functions at the Eulerian scale by applying the formalism recently derived in \cite{Doyon17,DoyonSphon17}.

This article is organized as follows. In Sec.~\ref{sec:LL_model} we introduce the Lieb-Liniger model, and review its Bethe ansatz solution. Our main result is summarized and discussed in Sec.~\ref{sec:main_results}, where the main formulas are presented. In Sec.~\ref{sec:shG} we introduce the sinh-Gordon field theory and review its non-relativistic limit. The derivation of our results is carried out in Sec.~\ref{sec:derivation},while Sec.~\ref{sec:applications} contains several applications. Our conclusions are gathered in Sec.~\ref{sec:conclusions}, while some technical aspects of our work are reported in a few appendices.

\section{The Lieb-Liniger model}\label{sec:LL_model}

We consider a one-dimensional gas of point-wise interacting bosons on a system of length $L$, described by the Lieb-Liniger Hamiltonian 
\be
H=\int_{0}^{L}\,{\rm d}x \left\{\frac{1}{2m}\partial_x\Psi^{\dagger}(x)\partial_x\Psi(x)+\kappa \Psi^{\dagger}(x)\Psi^{\dagger}(x)\Psi(x)\Psi(x)\right\}\,,
\label{eq:hamiltonian}
\ee
where periodic boundary conditions are assumed. Here $\Psi^{\dagger}(x)$,$\Psi(x)$ are bosonic creation and annihilation operators satisfying $\left[\Psi(x),\Psi^{\dagger}(y)\right]=\delta(x-y)$, while $\kappa>0$ is the interaction strength. 

The Hamiltonian \eqref{eq:hamiltonian} can be diagonalized by means of the Bethe ansatz. In analogy with the free case, to each eigenstate is associated a set of real parameters $\{\lambda_j\}_{j=1}^{N}$, called rapidities, which parametrize the corresponding wave function. The latter can be written down explicitly as
\bea
\psi_N\left(x_1,\ldots,x_N\right)&=& \sum_P A(P) \prod_{j=1}^N e^{i
	\lambda_{P_j} x_j}\,,\hspace{2pc} x_1\le x_2\le ...\le x_N\,,
\label{eq:wave_function}
\eea
where the sum is over all the permutations of the rapidities and the symmetric extension is assumed for a different ordering of the coordinates $\{x_j\}_{j=1}^N$.
The coefficients $A(P)$ are not independent and can be recursively obtained as follows. Denoting with $\Pi_{j,j+1}$ the permutation exchanging the rapidities at positions $j$ and $j+1$, we have
\be
A(\Pi_{j,j+1}P)=S_\text{LL}(\lambda_{P_j}-\lambda_{P_{j+1}})A(P)\, , 
\ee
where $S_{\rm LL}(\lambda)$ is the scattering matrix of the model
\be
S_{\rm LL}(\lambda)=\frac{\lambda-2im\kappa}{\lambda+2im\kappa}\,.
\ee
Physically, exchanging the order of rapidities in the Bethe wave function can be interpreted as a sequence of two-body scattering events. Imposing periodic boundary conditions results in a quantization of the rapidities, which is analogous to the free case. The presence of a non-trivial $S$-matrix, however, affects the quantization procedure, leading to the so-called Bethe equations 
\be
e^{i\lambda_j L}\prod_{k\neq j}^N S_{\rm LL}\left(\lambda_j-\lambda_k\right)=1\,.
\label{eq:bethe_eq}
\ee 
Given a solution to the system \eqref{eq:bethe_eq}, the momentum and energy of the corresponding eigenstate are immediately obtained as
\begin{equation}
P(\{\lambda_j\})=\sum_{j=1}^{N}p(\lambda_j)\ ,\qquad E(\{\lambda_j\})=\sum_{j=1}^{N}\epsilon(\lambda_j)\ ,
\label{momentum_energy}
\end{equation}
where $p(\lambda)=\lambda$ and $\epsilon(\lambda)=\lambda^2/(2m)$ are the single-particle momentum and energy respectively.
In addition to energy and momentum, the Lieb-Liniger Hamiltonian displays an infinite set of local conserved operators $\{Q_i\}$ \cite{DaKo11}. Their eigenvalues is still additive over the rapidities, namely
\be
\mathcal{Q}_i(\{\lambda_j\})=\sum_{j=1}^N \omega_i(\lambda_j)\, ,
\ee
where the state-independent functions $\omega_i$ are called the single-particle charge eigenvalues.

When the particle number grows to infinity, the rapidities associated to a given eigenstate arrange themselves on the real line according to a non-trivial distribution function $\rho(\lambda)$ \cite{takahashi-99}. In addition, one also introduces a hole distribution function $\rho_h(\lambda)$, which is analogous to the well-know distribution of unoccupied states for a free Fermi gas. In the thermodynamic limit, the Bethe equations \eqref{eq:bethe_eq} are translated into a constraint for the functions $\rho(\lambda)$ and $\rho_h(\lambda)$, which reads
\be
\rho(\lambda)+\rho_h(\lambda)=\frac{1}{2\pi}+\int_{-\infty}^{\infty}\frac{{\rm d}\mu}{2\pi}\,\varphi_\text{LL}(\lambda-\mu)\rho(\mu)\,,
\label{eq:thermo_bethe}
\ee
where
\be
\varphi_\text{LL}(\lambda)=-i\frac{\partial}{\partial\lambda}\log S_{\rm LL}(\lambda)=\frac{4m\kappa}{\lambda^2+4m^2\kappa ^2}\,.
\label{eq:kernelKappa}
\ee
In the following, we will omit the index LL when this does not generate confusion. The distribution function $\rho(\lambda)$ completely characterizes an eigenstate of the Hamiltonian in the thermodynamic limit: two states sharing the same rapidity distribution function are indistinguishable as far as the expectation values of local observables and their correlators are concerned. For example, the particle and energy densities are given by
\be
D=\int_{-\infty}^{\infty}{\rm d}\lambda\, \rho(\lambda)\,,\qquad e=\int_{-\infty}^{\infty}{\rm d}\lambda\,\epsilon(\lambda) \rho(\lambda)\,,\label{eq:density_energy}
\ee
while, more generally, expectation values of the local conserved quantities $Q_i=\int \dd x\,  q_i(x)$ read
\be
\langle q_i\rangle= \int^\infty_{-\infty} \dd \lambda\,  \omega_i(\lambda) \rho(\lambda)\, .
\ee

Note that the equation \eqref{eq:thermo_bethe} does not uniquely fix the function $\rho(\lambda)$, and additional constraints have to be imposed in order to identify the specific state under study. Besides single highly excited eigenstates, the rapidity distribution function can also describe suitable ensembles, e.g. the thermal ensembles and proper generalizations. Indeed, as far as expectation values of local operators are concerned, averaging on a given ensemble is completely equivalent to computing expectation values on a single, representative eigenstate: this is well known in the thermal case \cite{takahashi-99}, and has also been recently established more generally for GGEs, within the so-called Quench Action method \cite{CaEs13,Caux16}. In general, the representative eigenstate can be selected by evaluating the saddle-point of a suitable functional, which leads to an integral equation for the corresponding rapidity distribution functions. In addition to the Bethe equations \eqref{eq:thermo_bethe}, the latter uniquely fixes the macrostate. In order to exemplify this in the case of thermal states, we introduce 
\be
e^{\varepsilon(\lambda)}=\frac{\rho_h(\lambda)}{\rho(\lambda)}\,,\qquad \vartheta(\lambda)=\frac{1}{e^{\varepsilon(\lambda)}+1}\,,
\label{eq:filling}
\ee
where $\vartheta(\lambda)$ is usually referred to as the filling function.
Then, the integral equation characterizing the thermal representative eigenstate reads \cite{takahashi-99}
\be
\varepsilon(\lambda)=\beta\big[\epsilon(\lambda)-\mu\big]-\int_{-\infty}^{\infty}\frac{{\rm d}\mu}{2\pi}\varphi_{\text{LL}}(\lambda-\mu)\log\left(1+e^{-\varepsilon(\mu)}\right)\,.
\label{eq:thermal_TBA}
\ee
Here $\beta=1/T$ is the inverse temperature, while $\mu$ is a chemical potential. Eq.~\eqref{eq:thermal_TBA} can be easily solved numerically together with \eqref{eq:thermo_bethe}.

In the following, we will also consider different kinds of states, focusing in particular on GGEs, which generalize the usual thermal Gibbs ensembles, taking into account all the higher local and quasi-local conservation laws \cite{RDYO07,ViRi16,EsFa16}. It is now well accepted that these states describe the properties of the system at late times after it is taken out of equilibrium, for example by means of a quantum quench \cite{cc-06}. In general, each GGE will be described by an appropriate integral equation analogous to \eqref{eq:thermal_TBA}, namely
\be
\varepsilon(\lambda)=w(\lambda)-\int_{-\infty}^{\infty}\frac{{\rm d}\mu}{2\pi}\varphi_{\text{LL}}(\lambda-\mu)\log\left(1+e^{-\varepsilon(\mu)}\right)\,,
\label{eq:betaGGE}
\ee
with a driving term $w(\lambda)$ which keeps into account all the relevant conservation laws, besides energy and number of particles. It is interesting to note that in a few cases the solution to \eqref{eq:betaGGE} could be determined analytically \cite{DWBC14,DeCa14,PiCE16,Bucc16,PiCE16}.

In this work we are interested in the computation of one-point functions on arbitrary thermodynamic states characterized by the solution to suitable equations of the form \eqref{eq:betaGGE}. Denoting with $|\{\lambda_j\}\rangle$ the eigenstate corresponding to the set $\{\lambda_j\}$, we focus in particular on
\be
g_K=\frac{\mathcal{O}_K}{D^N}\,,
\ee
where $D$ is the density defined in \eqref{eq:density_energy}, while
\be
\mathcal{O}_K\equiv \langle\rho|\left(\Psi^{\dagger}(x)\right)^K\Psi^K(x)|\rho\rangle=\lim_{\rm th }\langle \{\lambda_j\} |\left(\Psi^{\dagger}(x)\right)^K\Psi^K(x)|\{\lambda_j\} \rangle\,,
\label{eq:expectation_value}
\ee
where $\{\lambda_j\}$ is a set of rapidities which corresponds to the distribution $\rho$ in the thermodynamic limit. Our goal consists in expressing the expectation value \eqref{eq:expectation_value} only in terms of the rapidity distribution function $\rho(\lambda)$.

Given the many-body wave function \eqref{eq:wave_function} one could in principle compute all local correlations for finite system sizes $L$.  However, this can be done in practice only for small values of $L$, due to the complicated structure of the wave functions. In fact, exploiting this representation, the computation of local correlations usually involves sums of an exponentially large number of terms, which makes the computation in the thermodynamic limit extremely hard. On the other hand, a more sophisticated approach, the algebraic Bethe ansatz, can be used to derive exact formulas at finite size which are suitable for analytic computation also in the thermodynamic limit. While this program has in fact been successfully followed in several cases  \cite{CaCa06,IzKR87,ItIK90,Slav90,KoKS97,KoMS11,SPCI12,Kozl14,PiCa15,DePa15}, we will purse a different approach, based on a non-relativistic limit of the sinh-Gordon field theory. In the next section we will present our main results, while the details of our derivation are postponed to the subsequent sections.

\section{Summary of our results} \label{sec:main_results}

Our main result is a general formula for the thermodynamic limit of one-point functions in the Lieb-Liniger model. Using the notations of the previous subsection, it reads
\be
\langle\rho|\left(\Psi^{\dagger}(x)\right)^K\Psi^K(x)|\rho\rangle=(K!)^2\left(m\kappa\right)^{K}\sum_{\sum_j jn_j=K}\prod_j \left[\frac{1}{n_j!}\left(\frac{\mathcal{B}_j}{2\pi m\kappa} \right)^{n_j}\right]\,.
\label{eq:main_corr}
\ee
Here, the sum is taken over all the possible integers $n_j\ge 1$ such that the constraint $\sum_{j=1}^\infty jn_j=K$ is satisfied; the coefficients $\mathcal{B}_j$ are defined as
\be\label{eq:B_def}
\mathcal{B}_j=\frac{1}{j}\int_{-\infty}^{+\infty} \dd\lambda\, \vartheta(\lambda)b_{2j-1}(\lambda)\,,
\ee
where the functions $b_j(\lambda)$ satisfy the following set of integral equations
\bea
b_{2n}(\lambda)&=&\int_{-\infty}^{+\infty} \frac{\dd\mu}{2\pi}\,\vartheta(\mu)\left\{ \varphi(\lambda-\mu)[b_{2n}(\mu)-b_{2n-2}(\mu)]
+\Gamma(\lambda-\mu)[2b_{2n-1}(\mu)-b_{2n-3}(\mu)]\right\}\,, 
\label{eq:int_1}
\eea
\bea
b_{2n+1}(\lambda)&=&\delta_{n,0}+\int_{-\infty}^{+\infty} \frac{\dd\mu}{2\pi}\, \vartheta(\mu)\big\{\Gamma(\lambda-\mu) b_{2n}(\mu)+ \varphi(\lambda-\mu)[b_{2n+1}(\mu)-b_{2n-1}(\mu)]\big\}\,,
\label{eq:int_2}
\eea
with the convention $b_{j\le 0}(\lambda)=0$ and
\be
\Gamma(\lambda)=\frac{2\lambda}{\lambda^2+(2 m\kappa)^2}\, .
\label{eq:kernelGamma}
\ee
Eq.~\eqref{eq:main_corr} is most easily encoded in the following generating function
\be\label{eq:generating_function}
1+\sum_{n=1}^{\infty}X^n\frac{\langle(\Psi^\dagger)^n(\Psi)^n\rangle}{(n!)^2(\kappa m)^n}=\exp\left(\frac{1}{2\pi m\kappa}\sum_{n=1}^\infty X^n \mathcal{B}_n\right)\,,
\ee
where we omitted the spatial dependence of the bosonic fields. Comparison between the Taylor expansion in $X$ on both sides gives \eqref{eq:main_corr}.

We note the hierarchical structure of the integral equations above. Indeed, each equation is a linear integral equation for a given unknown function $b_i(\lambda)$, where $b_{j<i}(\lambda)$ only contribute as source terms. In this perspective, obtaining the functions $b_{j\le 2K-1}(\lambda)$  (and thus the $K^\text{th}$ one point function in the Lieb Liniger gas), boils down to solving recursively $2K-1$ linear integral equations. The latter can be easily solved for example by a simple iterative scheme.

For the sake of clarity, we explicitly write down Eq.~\eqref{eq:main_corr} for the first values of $K$; in particular, up to $K=4$ we have
\bea
\mathcal{O}_2&=&\kappa^2\left(\frac{\mathcal{B}_1^2}{2\pi^2\kappa^2}+\frac{\mathcal{B}_2}{\pi\kappa}\right)\,,\\%
\mathcal{O}_3&=&\frac{36 \kappa^3}{8}\left(\frac{\mathcal{B}_1^3}{6\kappa^3\pi^3}+\frac{\mathcal{B}_1\mathcal{B}_2}{\pi^2\kappa^2}+\frac{\mathcal{B}_3}{\pi\kappa}\right)\,,\\
\mathcal{O}_4&=&36 \kappa ^4 \left(\frac{\mathcal{B}_1^4}{24 \pi ^4 \kappa ^4}+\frac{\mathcal{B}_2\mathcal{B}_1^2}{2 \pi ^3 \kappa ^3}+\frac{\mathcal{B}_3 \mathcal{B}_1}{\pi ^2 \kappa ^2}+\frac{\mathcal{B}_2^2}{2 \pi ^2 \kappa ^2}+\frac{\mathcal{B}_4}{\pi  \kappa }\right)\,.
\eea

\subsection{Discussion}

It is useful to compare our formulas with existing results in the literature. As we discussed in Sec.~\ref{sec:intro}, efficient integral formulas were already known for $\mathcal{O}_K$ with $K=2,3,4$ \cite{KoCI11,Pozs11}. These are also expressed in terms of the solution to simple integral equations, but their form differs from the one we found. It is non-trivial to see the equivalence between the two, which is most easily established numerically. In this respect, we extensively tested that our formulas give the same results of those of \cite{KoCI11,Pozs11} for $K=2,3,4$ and different macrostates. Furthermore, it is possible to show the equivalence by means of a perturbative analytical expansion in the filling function $\vartheta(\lambda)$. The calculations are rather technical, and are reported in Appendix~\ref{app:checks}.

For higher $K$ and before of our result, we could reside either on the LeClair-Mussardo expansion \cite{LeMu99}, or on the multiple integral representation derived in \cite{Pozs11}, the latter being equivalent to a resummation of the whole LeClair-Mussardo series. However, the presence of multiple integrals makes the computation of these expressions unfeasible on a practical level. On the other hand, analytic results for thermal states and generic $K$ were obtained in \cite{NRTG16} in the limit of large interactions; while predictions have been obtained also for non point-wise correlations, the latter are valid only for ground and thermal states.

In summary, not only our formulas provide an exact representation of the one-point correlations in arbitrary macrostates, but are also entirely expressed in terms of simple integrals of the solution to linear integral equations. This makes them particularly convenient for numerical evaluation, providing a full solution to the problem of computing one-point functions in the Lieb-Liniger model.

\section{The sinh-Gordon field theory and its non-relativistic limit}
\label{sec:shG}

In this section, we introduce the sinh-Gordon field theory, and briefly review its non-relativistic limit to the Lieb-Liniger model. In the following, we will only present the main aspects which will be relevant for our work, referring the reader to \cite{KoMT09,KoMT11} for a more detailed treatment.

The sinh-Gordon model is a quantum field theory of a real field $\phi$, whose action reads
\be
\mathcal{S}_\text{shG}=\int \dd x\dd t\,  \frac{1}{2c^2}(\partial_t\phi)^2-\frac{1}{2}(\partial_x\phi)^2-\frac{mc^4}{16\kappa }(\cosh(c^{-1}4\sqrt{\kappa}\phi)-1)\, .
\label{shGaction}
\ee
Note the unconventional choice of the notation, which has been chosen for later convenience. The integrability of the model is well known, both at the classical \cite{Grauel85} and at the quantum \cite{Dorey} level. The scattering matrix of the shG model was firstly computed in \cite{AFZ1983} and its analysis confirmed the presence of a single excitation species (see Ref. \cite{smirnov} or Ref. \cite{Mbook} for a complete discussion). Explicitly, it reads
\be
S_{\text{shG}}(\theta)=\frac{\sinh\theta-i\,\sin(\pi\alpha)}{\sinh\theta+i\,\sin(\pi\alpha)}\,\,\,,
\ee
where the parameter $\alpha$ is
\be
\alpha \,= \,\frac{c^{-1}16\kappa}{8\pi + c^{-1}16 \kappa} \,\,\,.
\ee

The study of thermodynamic properties of the shG model can be performed by means of the thermodynamic Bethe ansatz, and can be carried out in analogy with thermodynamic treatment of the Lieb-Liniger model. In particular, a given macrostate of the theory will be characterized by rapidity and hole distribution functions $\rho(\lambda)$ and $\rho_h(\lambda)$ which, in analogy to \eqref{eq:thermo_bethe}, will be constrained to satisfy some Bethe equations
\be
\rho(\lambda)+\rho_h(\lambda)=\frac{Mc\cosh\theta}{2\pi}+\int_{-\infty}^{\infty}\frac{{\rm d}\mu}{2\pi}\,\varphi_\text{shG}(\lambda-\mu)\rho(\mu)\,.
\label{eq:shGBethe}
\ee
These are identical to those of the Lieb Liniger model \eqref{eq:thermal_TBA}, provided the Galilean dispersion law is replaced with the relativistic one and that one uses the kernel $\varphi_\text{shG}$ derived from the shG $S-$matrix.
While the $S-$matrix is enough to describe the thermodynamics of the model, as well as expectation values of conserved charges, it does not provide other quantities of interests, such as the one point correlators of the bosonic field $\phi$ and its powers. In this case, extra information is needed, being the latter encoded in the form factors \cite{smirnov}.
Given a multi-particle state $\ket{\theta_1,...,\theta_n}$ and a local observable $\mathcal{O}(0,0)$ placed at $x=t=0$, the form factor $F^\mathcal{O}_n$ is the matrix element between the state and the vacuum
\be
F^\mathcal{O}_n(\theta_1,...,\theta_n)=\langle 0|\mathcal{O}|\theta_1,...,\theta_n\rangle\, .
\ee
More general matrix elements are obtained exploiting the crossing symmetry
\be\label{cross_form}
\langle \theta_1,...,\theta_n|\mathcal{O}(0,0)|\beta_1,...,\beta_m\rangle= F^\mathcal{O}_{n+m}(\beta_1,...,\beta_m,\theta_1-i\pi,...,\theta_n-i\pi)\, .
\ee
The form factors must obey several constraints known as Watson equations \cite{watson}
\be
F_n^\mathcal{O}(\theta_1,...,\theta_i,\theta_{i+1},...,\theta_n)=S_\text{shG}(\theta_i-\theta_{i+1})F_n^\mathcal{O}(\theta_1,...,\theta_{i+1},\theta_i,...,\theta_n)\, ,
\ee
\be
F_n^\mathcal{O}(\theta_1+2\pi i,\theta_2,...,\theta_n)=\prod_{i=2}^n S_\text{shG}(\theta_i-\theta_1) F_n^{\mathcal{O}}(\theta_1,\theta_2,...,\theta_n)\, ,
\ee
which guarantee the consistency of the form factors under interchanges of particles. The analytical structure of the form factors can be further understood by means of physical considerations; in fact a singularity is expected whenever two rapidities differ of $i\pi$. This singularity is associated with the annihilation process of a particle and an antiparticle, as it can be seen from \eqref{cross_form}. In particular, such a singularity is a pole whose residue is associated with the form factors with the annihilated particles removed \cite{smirnov}, namely
\be\label{boot_res}
-i\text{Res}_{\tilde{\theta}=\theta} F_{n+2}^\mathcal{O}(\tilde{\theta}+i\pi,\theta,\theta_1,...,\theta_n)=\left(1-\prod_{i=1}^n S_\text{shG}(\theta-\theta_i)\right) F_n^\mathcal{O}(\theta_1,...,\theta_n)\, .
\ee
The bootstrap program consists in trying to determine the form factors from these equations, together with some minimal assumptions. Such a program was successfully carried out in the shG case by Koubek and Mussardo \cite{KoubecMuss}, who obtained a closed expression for form factors of the vertex operators $e^{kc^{-1}4\sqrt{\kappa}\phi}$ for any $k$ on arbitrary states, which reads
\be
F^k_n=	\langle 0| e^{kc^{-1}4\sqrt{\kappa}\phi}|\theta_1,...,\theta_n\rangle =\frac{\sin(k c\pi \alpha)}{\pi\alpha}\Big(\frac{4\sin(\pi\alpha)}{N}\Big)^{n/2}\det M_n(k) \prod_{i<j}^n\frac{F_\text{min}(\theta_i-\theta_j)}{e^{\theta_i}+e^{\theta_j}}\,.
\ee
Here we introduced
\be\label{norm_cost}
N=\frac{1}{\cos(\pi\alpha/2)}\exp\Bigg[-\frac{1}{\pi}\int_0^{\pi\alpha}dt\frac{t}{\sin(t)}\Bigg]\, ,
\ee
\be\label{min_F}
F_\text{min}(\theta)=N\exp\Bigg[4\int_0^\infty \frac{dt}{t}\frac{\sinh(t\alpha/2)\sinh(t(1-\alpha)/2)}{\sinh(t)\cosh(t/2)}\sin^2\Big(\frac{t(i\pi-\theta)}{2\pi}\Big)\Bigg]\, .
\ee
Finally, the matrix $M_n(k)$ is defined as
\be
[M_n(k)]_{ij}=\sigma_{2i-j}^{(n)}\frac{\sin\left[\left((i-j)c^{-1}+k\right)c\pi\alpha\right]}{\pi\alpha}\, ,
\ee
where the indexes $i,j$ run from $1$ to $n-1$ and $\sigma_{i}^{(n)}$ are the symmetric polynomials defined as
\be\label{sym_pol}
\prod_{i=1}^n(x+e^{\theta_i})=\sum_{k=1}^n x^{n-k}\sigma_k^{(n)}\, .
\ee
As a simple, but crucial, byproduct the form factors of the powers of the field $\phi^n$ can also be obtained, by means of a simple Taylor expansion in $k$ of the vertex operators.

The form factors are the building blocks for the computation of local expectation values; in particular, in integrable field theories with diagonal scattering matrix, they enter directly into the so-called LeClair–Mussardo series \cite{LeMu99}. The latter is a remarkable tool for the computation of one-point functions, and within our notations reads
\be
\langle \mathcal{O} \rangle=\sum_{k=0}^\infty \frac{1}{k!}\int \frac{\dd^k \theta}{(2\pi)^k}\left(\prod_{j=1}^k \vartheta(\theta_j) \right)\langle\theta_k,...,\theta_1|\mathcal{O}|\theta_1,...,\theta_k\rangle_\text{c}\, .
\ee
Here, the connected matrix element is defined by a careful removal of the kinematical singularities \eqref{boot_res}
\be
\langle\theta_k,...,\theta_1|\mathcal{O}|\theta_1,...,\theta_k\rangle_\text{c}=\text{finite part}\left(\lim_{\epsilon_i\to 0}F^\mathcal{O}_{2n}(\theta_1,...,\theta_k,\theta_k-i\pi+i\epsilon_k,...,\theta_1-i\pi+i\epsilon_1)\right)\, ,
\ee
where the limit $\epsilon_i\to 0$ must be taken independently. 
This expansion was firstly conjectured in \cite{LeMu99}, verified on the set of the local charges in \cite{saleur2000} and finally rigorously proven for generic states in \cite{Pozs11_mv}. The LeClair-Mussardo series involves in general multiple coupled integrals that make impossible a straightforward resummation, even though in many cases its truncation to the first few terms provides quite accurate results.

More recently a remarkable expression, equivalent to a resummation of the LeClair-Mussardo series,  was achieved by Negro and Smirnov \cite{NeSm13,Negr14} for a particular class of vertex operators.
 The formula was later slightly simplified in Ref. \cite{BePC16}, where it was cast into the extremely simple form
\be
\frac{\langle e^{(k+1)c^{-1}4\sqrt{\kappa}\phi}\rangle}{\langle e^{kc^{-1}4\sqrt{\kappa}\phi}\rangle}=1+\frac{2\sin(\pi\alpha(2k+1))}{\pi}\int_{-\infty}^\infty \dd\theta \, \vartheta(\theta)e^\theta p_k(\theta)\label{BPeq}\, ,
\ee
where $k$ is a positive integer with $p_k(\theta)$ being the solution of the following integral equation
\be
p_k(\theta)=e^{-\theta}+\int_{-\infty}^\infty \dd\theta' \vartheta(\theta')\chi_k(\theta-\theta')p_k(\theta'),\hspace{2pc}
\chi_k(\theta)=\frac{i}{2\pi}\left(\frac{e^{-i2k\alpha\pi}}{\sinh(\theta+i\pi\alpha)}-\frac{e^{i2k\alpha\pi}}{\sinh(\theta-i\pi\alpha)}\right)\, .
\ee
Eq. \eqref{BPeq} gives us access to ratios of vertex operators, whose value can be iteratively computed. In fact, $e^{kc^{-1}4\sqrt{\kappa}\phi}$ for $k=0$ reduces to the identity operator, whose expectation value is trivially $1$: expectation values $\langle e^{kc^{-1}4\sqrt{\kappa}\phi}\rangle$ are then recovered for integers $k$ by mean of a repetitive use of eq. \eqref{BPeq}.
As pointed out in Ref. \cite{NeSm13,Negr14,BePC16}, arbitrary vertex operators are in principle obtainable thanks to a special symmetry of the sinh-Gordon model. 

In fact, by mean of an accurate analysis of the LeClair-Mussardo series and using the exact form factors, it is possible to show that $e^{kc^{-1}4\sqrt{\kappa}\phi}$  is periodic in $k$ with period $\alpha^{-1}$. Thus, if $\alpha$ is irrational (this is not a limitation, since any number is arbitrary well approximated by an irrational one) the sequence $\{k \mod \alpha^{-1}\}_{k=1}^\infty$ is dense in $[0,\alpha^{-1})$ and the expectation values of all the vertex operators are in principle recovered.

\subsection{The non-relativistic limit}

We now finally review the non-relativistic limit of the sinh-Gordon model. Here we provide only a short summary of the basic formulas; we refer the interested reader to the original paper \cite{KoMT09} for more details, as well as to \cite{KoCI11,KoMP10,KoMT11,CaKL14} where various applications have been presented (see also \cite{BaLM16,BaLM17} for non-relativistic limits of other integrable field theories).

The correspondence between the sinh-Gordon and the Lieb-Liniger models can be established by looking at the corresponding scattering matrices. In fact, by means of a comparison between the momentum eigenvalues, it is natural to set $\theta\sim \lambda/mc$ (where we recall $\lim_\text{NR}M= m$). Using this correspondence, it is immediate to realize
\be
\lim_{c\to\infty}S_\text{shG}(c^{-1}m^{-1}\lambda)=S_\text{LL}(\lambda).
\ee
The limit is immediately extended to the whole thermodynamics through the Bethe equations. In particular, the relation between the shG excitation distribution function and the Lieb-Liniger one can be easily understood looking at the total excitation density. In particular, requiring 
\be
D=\int_{-\infty}^\infty \dd\theta\, \rho^\text{shG}(\theta)=\int_{-\infty}^\infty \dd\lambda\, c^{-1}m^{-1}\rho^\text{shG}(c^{-1}m^{-1}\lambda)\,,
\ee
one is led to the natural identification 
\be
\rho^\text{LL}(\lambda)=c^{-1}m^{-1}\rho^\text{shG}(c^{-1}m^{-1}\lambda)\,,
\ee
where superscripts shG and LL are introduced to distinguish the two densities. This scaling guarantees that the Bethe equations for the shG model \eqref{eq:shGBethe} become those of the LL gas \eqref{eq:thermo_bethe}, provided the hole distribution is rescaled in the same way, namely $\rho_h^\text{LL}(\lambda)=c^{-1}m^{-1}\rho_h^\text{shG}(c^{-1}m^{-1}\lambda)$: in particular, this implies the filling function of the shG field theory reduces, at the leading order, to the filling function in the LL model $\vartheta^\text{shG}(c^{-1}m^{-1}\lambda)=\vartheta^\text{LL}(\lambda)$.

The non-relativistic limit is slightly more involved at the level of correlation functions. The starting point is provided by the following mode-splitting of the relativistic field
\be
\phi(t,x)=\frac{1}{\sqrt{2m}}\left(e^{imc^2t}\Psi^{\dagger}(t,x)+e^{-imc^2t}\Psi(t,x)\right)\label{LL11}\,\,\,.
\ee
The exponential oscillating terms are introduced to take care of the divergent $\sim m^2c^2$ contribution coming from the Taylor expansion of the interaction in the shG action \eqref{shGaction}. The fields $\Psi$, assumed to be smooth functions of space and time in the $c\to\infty$ limit, can be interpreted as the field operators in the Lieb-Liniger model. This claim is also supported by the analysis of the momentum conjugated to $\phi$, hereafter denoted as $\Pi$ and defined as
\be
\Pi(t,x)\,=\,\frac{1}{c^2}\partial_{t}\phi(t,x)\,=\,i\sqrt{\frac{m}{2}}\left(e^{imc^2t}\Psi^{\dagger}(t,x)-e^{-imc^2t}\Psi(t,x)\right)+\mathcal{O}(c^{-2})\,. 
\ee
Indeed, one can see that the relativistic commutation rules $[\phi(t,x),\Pi(t,y)]=i\delta(x-y)$ are in fact consistent with the non-relativistic relations
\be
[\Psi(t,x),\Psi(t,y)]=0,\hspace{3pc}[\Psi(t,x),\Psi^{\dagger}(t,y)]\,=\,\delta(x-y) \, . 
\ee
Dynamically, the correspondence between the shG and LL models is corroborated by the non-relativistic limit of the action. In fact, plugging \eqref{LL11} into the shG action \eqref{shGaction}, neglecting the vanishing terms together with the fast oscillating phases, we readily obtain
\be
\lim_{\text{NR}}\mathcal{S}^{\text{shG}}\,=\int \dd x\dd t\,\left\{ \frac{i}{2}\left(\partial_t\Psi^\dagger\Psi-\Psi^\dagger\partial_t\Psi\right)-\frac{1}{2m}\partial_x\Psi^\dagger\partial_x\Psi-\kappa \Psi^\dagger\Psi^\dagger\Psi\Psi\right\}\label{LL18} \,,
\ee
namely, the action for the Lieb-Liniger model. Establishing the limit at the level of action hides some dangerous pitfalls that can lead to erroneous results when applied to other integrable field theories (see Refs. \cite{BaLM16,BaLM17} for more details). Nevertheless, this procedure in the shG case is correct and leads to the correspondence
\be\label{op_NR}
\lim_\text{NR}\langle:\phi^{2K+1}:\rangle=0,\hspace{2pc} \lim_\text{NR}\langle:\phi^{2K}:\rangle=\binom{2K}{K}\frac{1}{(2m)^K}\langle(\Psi^\dagger)^K(\Psi)^K\rangle\, ,
\ee
where $: \,\, :$ stands for normal ordering. Note that the correspondence can be simply understood by plugging the mode expansion \eqref{LL11} into $\phi^{2K}$ and dropping the oscillating phases.

The normal ordering procedure requires further comments. Inserting the mode expansion \eqref{LL11} into $\phi^{2K}$ and using the commutation relations to obtain a normal ordered expression, we obtain two types of terms: the first one is $(\Psi^\dagger)^K(\Psi)^K$; the second consists of products of fields $(\Psi^\dagger)^n(\Psi)^n$ with $n< K$, coupled to UV-singular terms coming from equal point commutators $\delta(0)$.
Of course, the output of the LeClair-Mussardo series (as well as of the Negro-Smirnov formula) refers to the renormalized fields, where UV-divergent quantities have been removed. However, it remains true that all the normal ordered fields $:\phi^{2n}:$ with $n\le K$ contribute to the expectation value of $\phi^{2K}$. In order to obtain the Lieb-Liniger one point functions from the shG formulas presented in the previous section, the decomposition of $\phi^{2K}$ in normal ordered fields must be performed explicitly.

A consistent derivation is performed assuming a linear mixing between normal ordered and non-normal ordered fields \cite{KoMT09}
\be
:\phi^{2K}:=\phi^{2K}-\sum_{n=1}^{K-1}\left(\frac{\kappa}{4}\right)^{(K-j)}\mathcal{N}^{K}_{n}\phi^{2n}\label{normTay}\,.
\ee
Here the coefficients $\mathcal{N}^K_j$ can be fixed as follows. The field $\phi^{2K}$ usually has non trivial matrix elements in each particle sector, namely $\langle 0|\phi^{2K}|\theta_1,...,\theta_{n'}\rangle\ne 0$. Instead, $:\phi^{2K}:$ is required to have trivial matrix element between the vacuum and the whole $n< 2K$ particle sector
\be\label{norm_def}
\langle 0|:\phi^{2K}:|\theta_1,...,\theta_{n}\rangle=0, \hspace{2pc} \forall \, n<2K\, .
\ee
Imposing this condition and employing the exact form factors of the powers of the fields, one can derive the mixing coefficients $\mathcal{N}_n^K$. Being ultimately interested in the implications for the Lieb-Liniger model, we will work under the assumption of the NR limit $c\to\infty$, which allows us to simplify our calculations: for example, the normalization constant $N$ \eqref{norm_cost} simply becomes $1$. The symmetric polynomials $\sigma_k^{(n)}$ \eqref{sym_pol} hugely simplify as well, leading to the compact result \cite{CaKL14}
\be
\det M_n(k)\to\left(\frac{\sin(2k \kappa)}{c^{-1}2\kappa}\right)^{n-1} \det \Bigg[\binom{n}{2i-j} \Bigg]=\left(\frac{\sin(2k \kappa)}{c^{-1}2\kappa}\right)^{n-1}  2^{n(n-1)/2}\, .
\ee
Furthermore, we obtain
\be
F_\text{min}\Big(c^{-1}m^{-1}\lambda\Big)\to \frac{\lambda}{\lambda+i2m\kappa }\,,
\ee
where $F_\text{min}(\theta)$ is defined in \eqref{min_F} .
Putting these terms together we obtain the non-relativistic limit of the form factor of the vertex operators, which reads
\be
F_n^k\to c^{n/2} 2^{n(n-1)/2}\left(\frac{\sin(2k\kappa)}{ \sqrt{2\kappa}}\right)^{n}\prod_{i<j}^n\frac{\lambda_{i}-\lambda_j}{\lambda_{i}-\lambda_j+i2m\kappa }\label{NRFFvertex}\, .
\ee
The form factors of the powers of the fields can be simply obtained by means of a Taylor expansion in $k$ of the form factors of the exponential fields. Imposing \eqref{norm_def} on \eqref{normTay} immediately leads to the following constraint
\be\label{N_definition}
\sum_{j=1}^{K-1}\mathcal{M}_{a,j}\, \mathcal{N}^{K}_j=\mathcal{M}_{a,K},\hspace{2pc} a\in\{1,...,K-1\}, \hspace{2pc}\text{with}\,\,\, \mathcal{M}_{n,n'}=\partial_x^{2n'} \sin^{2n}(x)\Big |_{x=0}\, .
\ee
Remarkably, in the NR limit the normal ordering expression can be explicitly solved and the fields $\phi^{2K}$ expressed in terms of the normal ordered ones
\be\label{inv_norm}
\phi^{2K}=\sum_{j=1}^{K} \frac{\mathcal{M}_{j,K}}{(2j)!}\left(\frac{\kappa}{4}\right)^{K-j}:\phi^{2j}:\, .
\ee

The normal ordering procedure acquires a very simple form when applied to the vertex operators. Indeed, we have 
\be\label{eq:nr_vertex}
\lim_\text{NR}\langle e^{4q\sqrt{\kappa} \phi}\rangle=1+\sum_{j=1}^\infty \frac{(4q\sqrt{\kappa})^{2j}}{(2j)!}\lim_\text{NR}\langle\phi^{2j}\rangle=1+ \sum_{n=1}^{\infty}\Bigg(\sum_{j=n}^\infty \frac{(4q\sqrt{\kappa})^{2j}}{(2j)!}\mathcal{M}_{n,j}\left(\frac{\kappa}{4}\right)^{j}\Bigg)\left(\frac{\kappa}{4}\right)^{-n}\lim_\text{NR}\langle:\phi^{2n}:\rangle\, .
\ee
Here $q$ is kept constant in the NR limit (the choice of such a normalization will be clear in the next section) and we made use of Eq.~\eqref{inv_norm}. Now, notice that
\be
\sum_{j=n}^\infty \frac{(4q\sqrt{\kappa})^{2j}}{(2j)!} \left(\frac{\kappa}{4}\right)^{j}\mathcal{M}_{n,j}=\frac{1}{(2n)!}\sum_{j=n}^\infty \frac{(4q\sqrt{\kappa})^{2j}}{(2j)!} \left(\frac{\kappa}{4}\right)^{j}\partial_x^{2j} \sin^{2n}(x)\Big |_{x=0}=\frac{\sin^{2n}(2q\kappa)}{(2n)!}\,,
\ee
which allows us to rewrite the NR limit of the vertex operator in the simple form
\be
\lim_\text{NR}\langle e^{4q\sqrt{\kappa} \phi}\rangle= \lim_\text{NR}\Big\langle  :e^{\frac{2}{\sqrt{\kappa}}\sin(2q\kappa)\phi}:\Big\rangle=1+ \sum_{n=1}^{\infty}\frac{\sin^{2n}(2q\kappa)}{(2n)!}\left(\frac{\kappa}{4}\right)^{-n}\lim_\text{NR}\langle:\phi^{2n}:\rangle\, .
\ee
By means of Eq.~\eqref{op_NR}, we can finally establish the following relation between the NR limit of vertex operators and the one point functions in the Lieb-Liniger model

\be\label{eq:nr_expansion}
\lim_\text{NR}\langle e^{4q\sqrt{\kappa}\phi}\rangle=1+ \sum_{n=1}^{\infty}\Big(1-\cos(4q\kappa)\Big)^n\frac{\langle(\Psi^\dagger)^n(\Psi)^n\rangle}{(n!)^2(m\kappa)^n}\, .
\ee

The next section is devoted to computing the NR limit of the vertex operator within the Negro-Smirnov formalism, concluding the derivation of our main result.

\section{One-point functions in the Lieb-Liniger model}
\label{sec:derivation}
The starting point for the derivation of our main result \eqref{eq:main_corr} is the Negro-Smirnov formula \eqref{BPeq}. A limit $c\to \infty$ with $k$ fixed of the l.h.s. leads to a trivial result. We will then follow the approach consisting in rescaling $k\to q=ck$ and subsequently take $c\to\infty$, while keeping $q$ fixed. In this case, at first order in $c^{-1}$ we obtain
\be
\frac{\langle e^{(q+c^{-1})4\sqrt{\kappa}\phi}\rangle}{\langle e^{q4\sqrt{\kappa}\phi}\rangle}=1+c^{-1}4\sqrt{\kappa}\lim_{c\to\infty}\left[\frac{\langle \phi e^{4q\sqrt{\kappa}\phi}\rangle}{\langle e^{4q\sqrt{\kappa}\phi}\rangle}\right]+...=1+c^{-1}\partial_q\lim_{c\to\infty}\partial_q \log\langle e^{4q\sqrt{\kappa}\phi}\rangle+...\,,
\ee
where the neglected terms are higher order in the $c^{-1}$ expansion. Note that the zeroth-order term is naturally canceled out by the r.h.s. of \eqref{BPeq} and the Negro-Smirnov formula reduces to
\be\label{NR_smirnov}
\partial_{q}\lim_\text{NR}\log\langle e^{4q\sqrt{\kappa}\phi}\rangle=\frac{2}{m\pi}\int_{-\infty}^\infty \dd\lambda\, \vartheta(\lambda)p^\text{LL}_q(\lambda)\, ,
\ee
where we defined
\be
p^\text{LL}_q(\lambda)=\lim_{c\to\infty}\left[\sin(4q\kappa) p_{cq}(c^{-1}m^{-1}\lambda)\right]\, .
\ee

From the NR limit of the integral equation satisfied by $p_k(\theta)$, we easily obtain an integral equation for $p_q^\text{LL}(\lambda)$
\be\label{int_pLL}
p_q^\text{LL}(\lambda)=\sin(4q\kappa)+\int_{-\infty}^\infty \dd\lambda' \,\vartheta(\lambda')\chi_q^\text{LL}(\lambda-\lambda')p_q^\text{LL}(\lambda'),\hspace{2pc}
\chi_q^\text{LL}(\lambda)=\frac{i}{2\pi}\left(\frac{e^{-iq4\kappa}}{\lambda+i 2m\kappa}-\frac{e^{iq4\kappa}}{\lambda-i2 m\kappa}\right)\, .
\ee
Thanks to the fact that $\langle e^{4q\sqrt{\kappa}}\rangle=1$ for $q=0$, we can explicitly integrate  eq. \eqref{NR_smirnov} and obtain an expression for the NR limit of the vertex operator
\be\label{eq:integrated_smirnov}
\lim_\text{NR}\langle e^{4q\sqrt{\kappa}\phi}\rangle=\exp\left[\frac{2}{m\pi}\int_{-\infty}^\infty \dd\lambda\, \vartheta(\lambda)\int_0^q \dd q'\,p^\text{LL}_{q'}(\lambda)\right]\, .
\ee
Looking at Eq. \eqref{eq:nr_expansion}, we can immediately understand that a convenient expansion of the above relation in terms of the trigonometric functions $\sin(4\kappa q)$ and $\cos(4\kappa q)$ will ultimately allow us to reach the one point functions in the LL model. In this perspective, we rewrite the kernel $\chi_q^\text{LL}(\lambda)$ as
\be
\chi_q^\text{LL}(\lambda)=\frac{1}{2\pi}\left[\cos(q4\kappa)\varphi_\text{LL}(\lambda)+\sin(q4\kappa)\Gamma(\lambda)\right]\,.
\ee
As it should be clear, an iterative solution to Eq.~\eqref{int_pLL} will naturally provide a power expansion in terms of the trigonometric functions $\sin(4q\kappa)$ and $\cos(4q\kappa)$. However, it is convenient to consider a different form of series expansion, and define the functions $A_j(\lambda)$ and $B_j(\lambda)$ as the coefficients of the series
\be
\int_0^q \dd q' \, p^\text{LL}_{q'}=\frac{1}{4\kappa}\sum_{j=0}^\infty \sin(q\kappa 4)(1-\cos(4\kappa q))^jA_j(\lambda)+\frac{1}{4\kappa}\sum_{j=1}^\infty(1-\cos(4q\kappa))^j B_j(\lambda)\,.
\label{eq:expansion}
\ee
For the moment, the functions $A_j(\lambda)$ and $B_j(\lambda)$ need to be determined. 
The form of this series is completely general and describes an arbitrary power series in terms of the trigonometric functions  $\sin(4\kappa q)$ and $\cos(4\kappa q)$. 
Taking the derivative with respect to $q$ of both sides of this equation we get
\be
 p^\text{LL}_{q}(\lambda)=\sum_{j=0}^\infty (1-\cos(4q\kappa))^jb_{2j}(\lambda)+\sin(q4\kappa)(1-\cos(4q\kappa))^jb_{2j+1}(\lambda)\,,
\label{eq:p_exp}
\ee
where
\bea\label{b_def}
b_{2j}(\lambda)=(2j+1)A_j(\lambda)-jA_{j-1}(\lambda)\,,\hspace{6pc}
 b_{2j+1}(\lambda)=(j+1)B_{j+1}(\lambda)\,.
\eea
The set of integral equations satisfied by $b_j(\lambda)$ are readily obtained using Eq.~\eqref{eq:p_exp} in the integral equation \eqref{int_pLL}. The derivation is long but straightforward, and leads to the set of integral equations \eqref{eq:int_1} and \eqref{eq:int_2}. Defining
\bea
\mathcal{A}_j&=&\int_{-\infty}^\infty \dd\lambda\, \vartheta(\lambda)A_j(\lambda)\,,\label{eq:aj_coefficient}\\ \mathcal{B}_j&=&\int_{-\infty}^\infty \dd\lambda\, \vartheta(\lambda)B_j(\lambda)\,,\label{eq:bj_coefficient}
\eea
and combining Eq.~\eqref{eq:nr_expansion} with Eq.~\eqref{eq:integrated_smirnov} we finally get 
\begin{eqnarray}
\nonumber&&1+ \sum_{n=1}^{\infty}\Big(1-\cos(4q\kappa)\Big)^n\frac{\langle(\Psi^\dagger)^n(\Psi)^n\rangle}{(n!)^2(m\kappa)^n}=\\
&&\exp\left(\frac{1}{2\pi m\kappa}\sum_{j=0}^\infty \sin(q\kappa 4)(1-\cos(4\kappa q))^j\mathcal{A}_j+\frac{1}{2\pi m\kappa}\sum_{j=1}^\infty(1-\cos(4q\kappa))^j \mathcal{B}_j\right)\,.
\end{eqnarray}

For consistency, we must have $\mathcal{A}_j=0$. This can be seen analytically as explained in Appendix~\ref{app:checkA}, as well as numerically to high precision for different filling functions $\vartheta(\lambda)$. Setting $\mathcal{A}_j=0$ and replacing $X=1-\cos(4q\kappa)$, we finally obtain \eqref{eq:generating_function}, and hence our main result Eq.~\eqref{eq:main_corr}.

\section{Applications}
\label{sec:applications}

In this section we present several applications of our main result \eqref{eq:main_corr}. In particular, after explicitly evaluating one-point functions for different macrostates, we discuss in detail the connection between the one point correlation functions and the full counting statistics of the particle number. Finally, we combine our result with the recently introduced generalized hydrodynamics \cite{CaDY16,BCDF16} to analyze inhomogeneous out-of-equilibrium protocols as well as correlation functions at the Eulerian scale \cite{DoyonSphon17,Doyon17}.

\subsection{Thermal states and global quenches}

\begin{figure}[t!]
\includegraphics[width=0.45\textwidth]{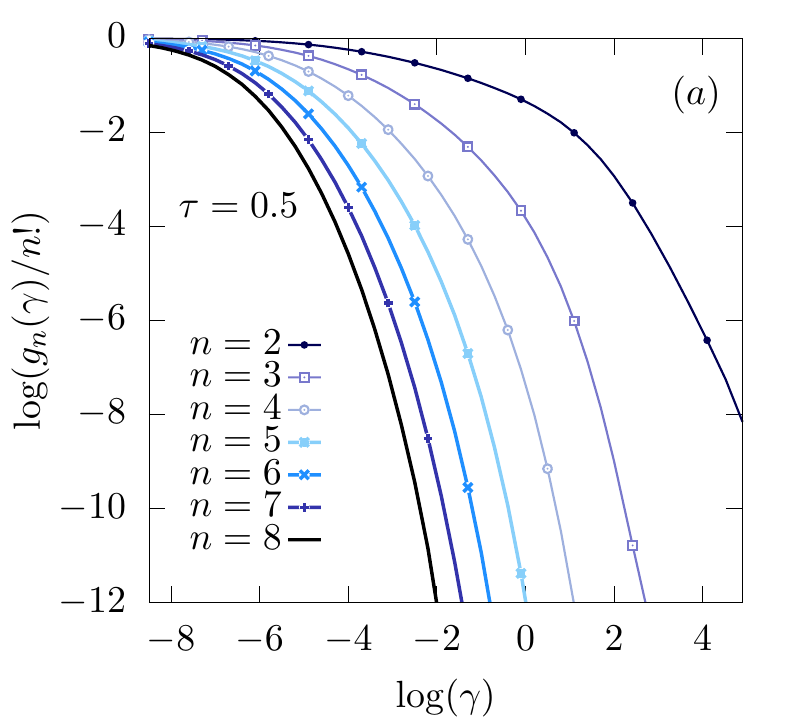}\hspace{2pc}
\includegraphics[width=0.45\textwidth]{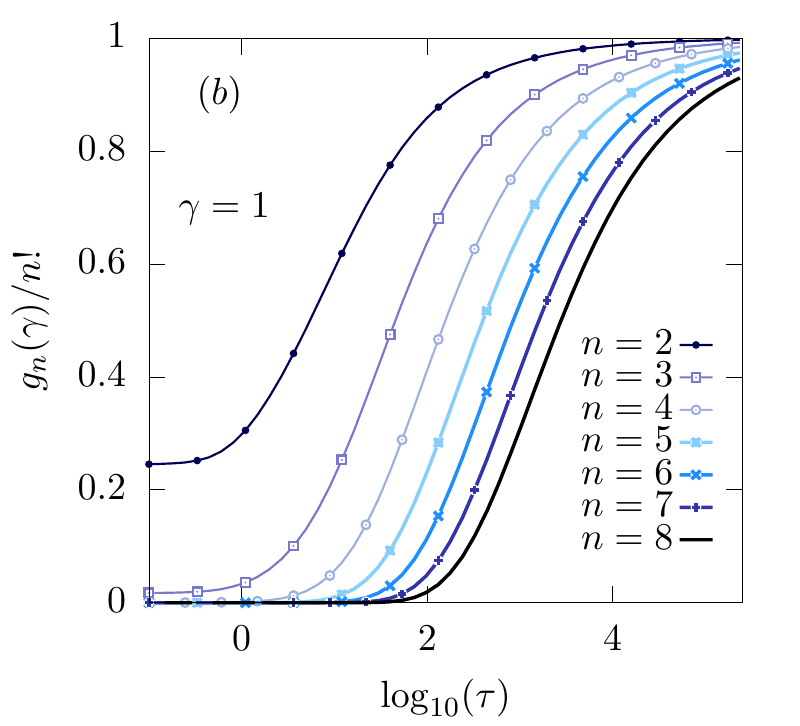}
\caption{The plots show the correlators $g_n$ computed using Eq.~\eqref{eq:main_corr} for thermal states $\rho=e^{-\beta H}/{\rm tr}[e^{-\beta H}]$. Subfigures $(a)$ and $(b)$ show the correlators as a function of the interaction $\gamma$ and the normalized temperature $\tau$ respectively (cf. the main text).}
\label{fig:thermal_states}
\end{figure}

In order to show the versatility of our formula \eqref{eq:main_corr}, we report its explicit evaluation for different macrostates. In Fig.~\ref{fig:thermal_states} we report explicit values of
the correlations for thermal states $\rho=e^{-\beta H}/{\rm tr}[e^{-\beta H}]$ as a function of the interaction [subfigure $(a)$] and of the temperature [subfigure $(b)$]. As a point of principle, we evaluated our formulas up to $n=8$ for a wide range of the parameters, showing that they are extremely suitable for numerical evaluation. Note that the correlators $g_n$ are only a function of the
rescaled parameters $\gamma=\kappa/D$ and $\tau=\beta^{-1}D^{-2}$. As a non-trivial check of our formulas we see that the limit $\lim_{\gamma\to 0}g_n(\gamma)=n!$ is recovered from our numerical results \cite{Pozs11}.  From subfigure $(a)$ of Fig.~\ref{fig:thermal_states}, it is apparent that $g_n(\gamma)$ vanishes for $\gamma \to \infty$ as it should. Furthermore, we verified that the decay at large $\gamma$ is algebraic, consistently with previous analytic findings in the literature \cite{NRTG16}. Analogously, it is possible to see from subfigure $(b)$ that $\lim_{\tau\to\infty} g_n=n!$; namely $g_n$ displays, for generic $n$, the same behavior of $g_{2}$, $g_{3}$ and $g_{4}$ \cite{Pozs11}. Finally, we see that $\lim_{\tau\to 0}g_{n}$ is a finite non-zero value (which depends on the interaction $\gamma$). 
 
\begin{figure}
	\includegraphics[width=0.45\textwidth]{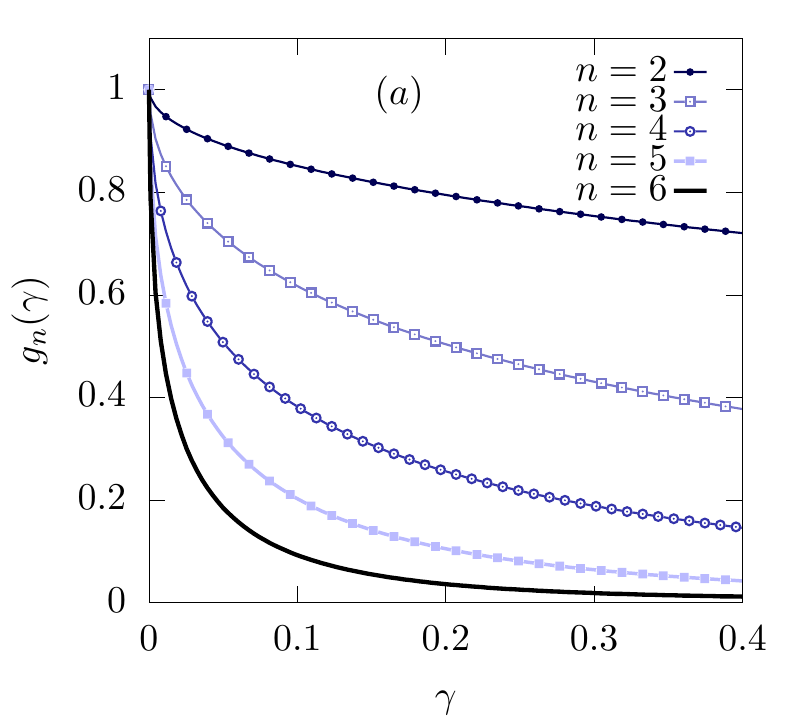}\hspace{2pc}
	\includegraphics[width=0.45\textwidth]{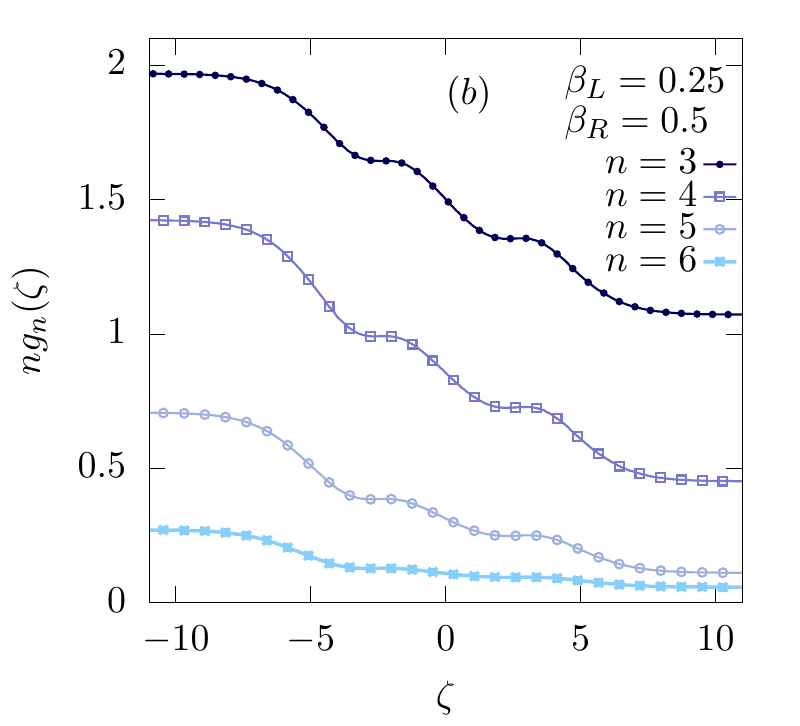}
	\caption{Local correlators on non-thermal states. Subfigure $(a)$: the plot shows the correlators $g_n$ computed using Eq.~\eqref{eq:main_corr} on the steady state reached at long times after an interaction quench where the initial state is the ground state of the non-interacting Hamiltonian \cite{DWBC14}. Subfigure $(b)$: Profiles of $g_n(\zeta)$ for the partitioning protocols studied in Refs.\cite{BCDF16,CaDY16}, cf. the main text. The two halves of the infinite systems, which are joined together at $t=0$ are prepared in thermals states with inverse temperatures $\beta_L=0.25$,$\beta_R=0.5$. The interaction coupling is fixed to be $\kappa=1$ ( the mass is set as usual to $m=1/2$).}
	\label{fig:quench_ness}
\end{figure}

As another example, we evaluated our formulas in two other physical situations. The first one is an interaction quench where the initial state is the ground state of the non-interacting Hamiltonian \cite{DWBC14}; at large time the system reaches a steady state whose rapidity distribution functions were computed analytically in \cite{DWBC14}, allowing us to obtain the corresponding local correlators. The latter are reported in subfigure $(a)$ of Fig.~\ref{fig:quench_ness}. Note in particular the different limiting behavior $\lim_{\gamma\to 0}g_n(\gamma)=1$.  Here, we still have a power-law decay at large values of $\gamma$: once again, the qualitative behavior of $g_n$ for general $n$ is the same of $g_2$ and $g_3$ computed in \cite{DWBC14}. The second physical situation that we consider is obtained by considering two halves of an infinite system which are prepared in two thermal states $\rho=e^{-\beta_{L/R} H}/{\rm tr}[e^{-\beta_{L/R} H}]$ with $\beta_L=1$, $\beta_L=2$ and suddenly joined together. At large time $t$ and distances $x$ from the junction, time- and space-dependent quasi-stationary states will emerge \cite{BCDF16,CaDY16}. In particular, a local relaxation to a GGE will occur for each ``ray'' $\zeta=x/t$, so that local observable will display non-trivial profiles as a function of $\zeta$ \cite{BCDF16,CaDY16}. We refer to Sec.~\ref{sec:hydro} for more details, while here we simply report in subfigure $(b)$ of Fig.~\ref{fig:quench_ness} an example of profiles for $\beta_L=0.25$ and $\beta_R=0.5$. Altogether, Figs.~\ref{fig:thermal_states} and \ref{fig:quench_ness} show unambiguously the great versatility of our formulas, which can be easily evaluated for very different physical situations. 

\subsection{The full counting statistics}
\label{sec:full_counting}

As one of the most interesting applications of our formulas, the knowledge of the expectation values of the one point functions $\langle (\Psi^\dagger)^K(\Psi)^K\rangle$ gives us access to the full counting statistics \cite{HLSI08,KPIS10,KISD11,GKLK12} of the number of particles within a small interval, as we show in this section. Given an interval of width $\Delta$, the mean number of particles we can measure in it is simply $\Delta \langle \Psi^\dagger\Psi\rangle$. However, the number of particles is a stocastic variable subjected to statistical fluctuations, and a full description of the quantum system should include the whole probability distribution of the latter, not only its mean value.

We define $\hat{N}_\Delta$ the operator which counts the number of particles within a small interval of length $\Delta$. In second quantization, it reads
\be
\hat{N}_\Delta=\int_0^\Delta \dd x \, \Psi^\dagger(x)\Psi(x)\, .
\ee
Its spectrum include all and only positive integers number, being its eigenvalues the number of particles. For this reason, we have the spectral decomposition
\be
\hat{N}_\Delta=\sum_{n=0}^\infty n \hat{P}_n\,,
\ee
where $\hat{P}_n$ is the projector on the space of fixed number $n$ of particles. Therefore, the probability of finding $n$ particles in the interval $n$ for a given macrostate $|\rho \rangle $ is the expectation value $P_{\Delta}(n)=\langle\rho|\hat{P}_n|\rho\rangle$, which is the object we aim to compute. In this respect, our main result is
\be\label{eq:full_count}
\lim_{\Delta\to 0}\frac{P_\Delta(n)}{\Delta^n}=\frac{\langle (\Psi^\dagger)^n( \Psi)^n \rangle}{n!}\,,
\ee
which will be derived in the following.

First, it is convenient to look at the generating function $\chi(\gamma)=\langle e^{i\gamma\hat{N}_\Delta}\rangle$. Indeed, $P_\Delta(n)$ is readily recovered from its Fourier transform
\be\label{eq:four_gen}
\int_{-\infty}^\infty \frac{\dd \gamma}{2\pi}\, e^{-i\gamma n'}\Big\langle e^{i\gamma \hat{N}_\Delta}\Big\rangle=\int_{-\infty}^\infty \frac{\dd \gamma}{2\pi}\, \Big\langle\sum_{n=0}^\infty e^{i \gamma(n-n')}P_n\Big\rangle=\delta(n-n')P_\Delta(n)\,.
\ee
It is useful to express $\chi(\gamma)$ in terms of normal ordered correlation functions. This can be achieved thanks to the following identity
\be\label{eq:norm_ord_full}
\exp\left[i\gamma\int_0^\Delta \dd x\, \Psi^\dagger(x) \Psi(x)\right]=:\exp\left[\left(e^{i\gamma}-1\right)\int_0^\Delta \dd x\, \Psi^\dagger(x) \Psi(x)\right]:\, ,
\ee
whose derivation is left to Appendix \ref{app:normal_order}. Making use of a power expansion of the normal ordered exponential, we obtain
\begin{eqnarray}
\nonumber&&\int \frac{\dd\gamma}{2\pi} \, e^{-i\gamma n}\Big\langle e^{i\gamma\int_0^\Delta\dd x\,\Psi^\dagger(x)\Psi(x)}\Big\rangle=\sum_{j=0}^\infty\frac{1}{j!}\int \frac{\dd\gamma}{2\pi} e^{-in\gamma}\left(e^{i\gamma}-1\right)^j \langle:\left(\int_0^\Delta \dd x\, \Psi^\dagger(x) \Psi(x)\right)^j :\rangle=\\
&&=\sum_{j=0}^\infty\frac{1}{j!}\sum_{m=0}^j\binom{j}{m}(-1)^{j-m}\left[\int \frac{\dd\gamma}{2\pi}\, e^{i\gamma (m-n)}\right] \langle:\left(\int_0^\Delta \dd x\, \Psi^\dagger(x) \Psi(x)\right)^j :\rangle\, .
\end{eqnarray}
In each term of the series expansion, the integration in $\gamma$ provides Dirac $\delta$s that constrain the support on integers values. Through a proper reorganization of the sum, we arrive at the final result
\be
P_\Delta(n)=\frac{1}{n!}\left[\sum_{j=0}^\infty\frac{(-1)^{j}}{j!}  \Big\langle:\left(\int_0^\Delta \dd x\, \Psi^\dagger(x) \Psi(x)\right)^{j+n} :\Big\rangle\right]\,.
\label{eq:full_fullcounting}
\ee
As it is clear, the one point functions do not determine the full counting statistics for arbitrary $\Delta$ and the whole multi-point correlators are needed. Nevertheless, in the $\Delta\to 0$ limit we can invoke the continuity of the correlators and extract the leading orders. We finally obtain
\be\label{eq:small_fullcounting}
P_\Delta(n)=\frac{1}{n!}\Delta^n \Big(\langle (\Psi^\dagger(0))^n (\Psi(0))^n\rangle+\mathcal{O}(\Delta)\Big)\,,
\ee
from which Eq. \eqref{eq:full_count} immediately follows.
The approximation which led from Eq. \eqref{eq:full_fullcounting} to Eq. \eqref{eq:small_fullcounting} is clearly valid if we can truncate the series, which requires the interval to be small if compared with the density $\Delta\ll D^{-1}$; furthermore, we assumed that the correlation functions are approximately constant on a range $\Delta$. This last condition can be estimated as $\Delta\ll \sqrt{D/\langle \partial_x\Psi^\dagger\partial_x\Psi\rangle}$.

\begin{figure}[t!]
	\begin{tabular}{ll}
	\includegraphics[width=0.45\textwidth]{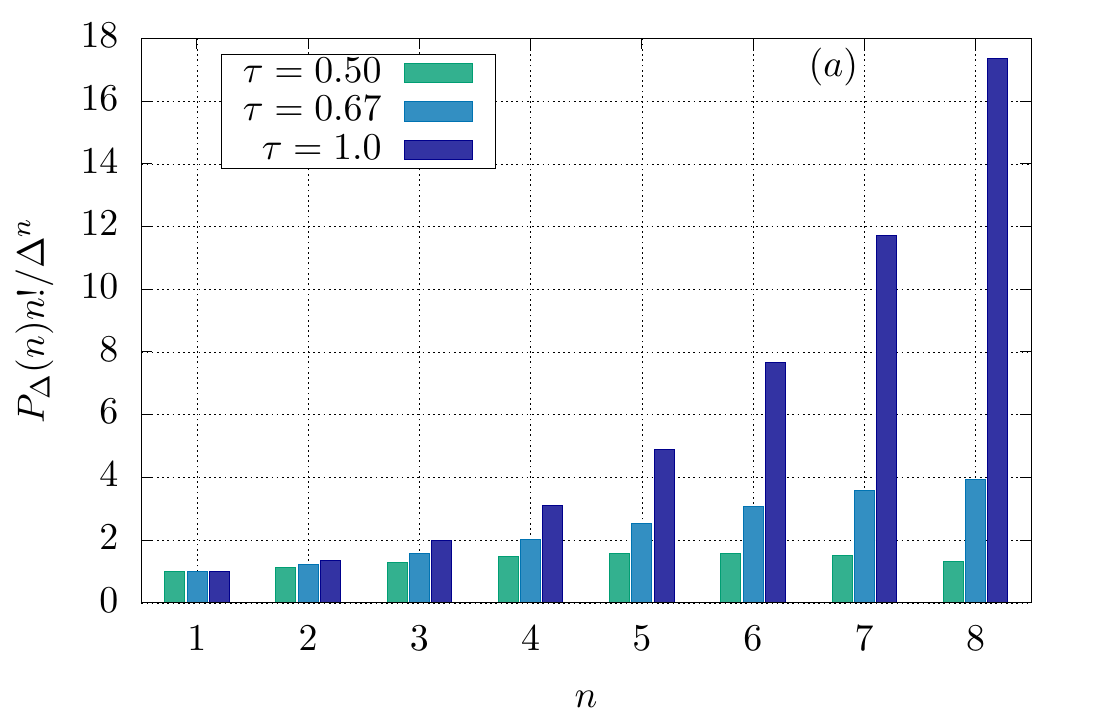} & 
	\hspace{0.25cm}\includegraphics[width=0.45\textwidth]{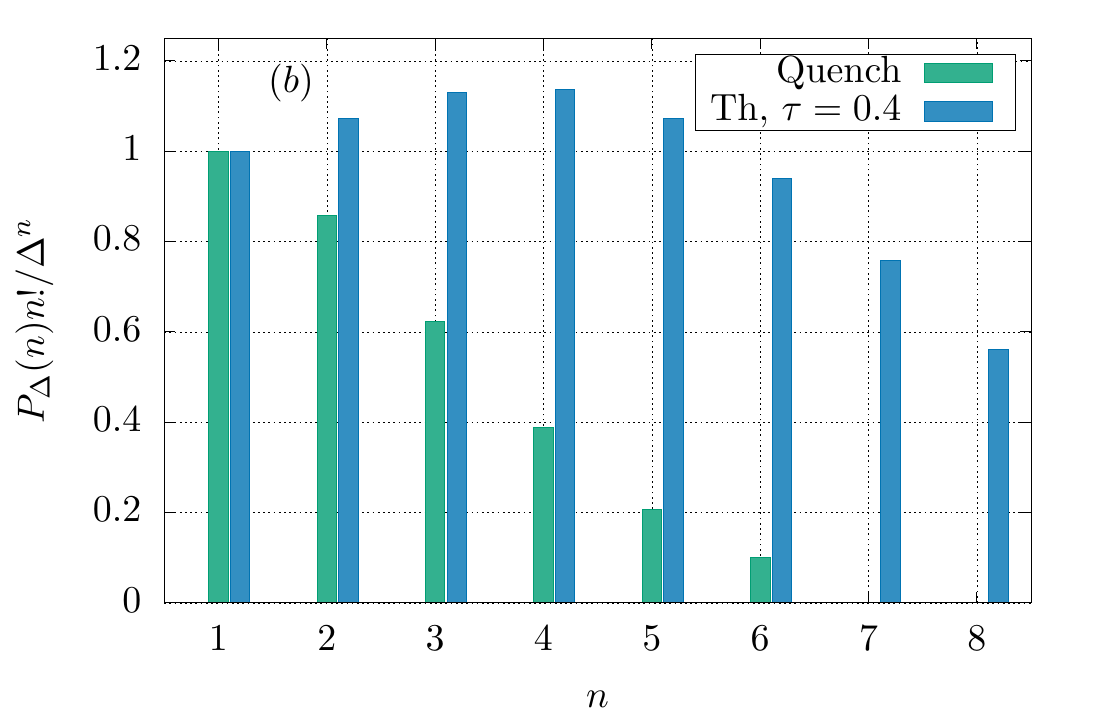}
\end{tabular}
\caption{Rescaled probability distribution for the particle number within an interval of length $\Delta$. Subfigure $(a)$: the plot shows results for different thermal states. The rescaled interaction is set to $\gamma=0.1$. Subfigure $(b)$: Comparison between a thermal state (with temperature $\tau =0.4$) and the post-quench steady state studied in \cite{DWBC14}.}
\label{fig:full_counting}
\end{figure}

From evaluation of Eq.~\eqref{eq:full_count}, it is clear that different macrostates display very different full counting statistics for the particle fluctuations. In particular, the latter provides a lot of information of a given macrostate. For the sake of presentation, we report in Fig.~\ref{fig:full_counting} the probabilities $P_\Delta(n)$ for different thermal states up to $n=8$. In subfigure $(a)$ we report results for thermal states at different values of the temperature $\tau$, and fixed interaction $\gamma$. We see that the magnitude of the normalized probabilities might vary significantly with the temperature. Furthermore, the behavior of $P_{\Delta}(n)n!$ is in general non-monotonic in $n$. This is even more manifest from subfigure $(b)$ of Fig.~\ref{fig:full_counting}, where we also report a comparison with the case of the post-quench steady state studied in \cite{DWBC14}.  Note that, in contrast, in this case $P_{\Delta}(n)n!$ displays a clear monotonic behavior. We stress that in these plots we restricted to small values of the interaction (here we chose $\gamma=0.1$) because in this case the values of $g_n(\gamma)$ (and hence of $P_{\Delta}(n)$) are larger: indeed, as it can be inferred from Fig.~\ref{fig:thermal_states} the value of $P_{\Delta}(n)$ decreases quickly as $\gamma$ increases. Altogether, these plots show the strong qualitative dependence of $P_{\Delta}(n)$ on the specific initial state considered.

\subsection{Hydrodynamics}
\label{sec:hydro}

\begin{figure}
	\includegraphics[width=0.6\textwidth]{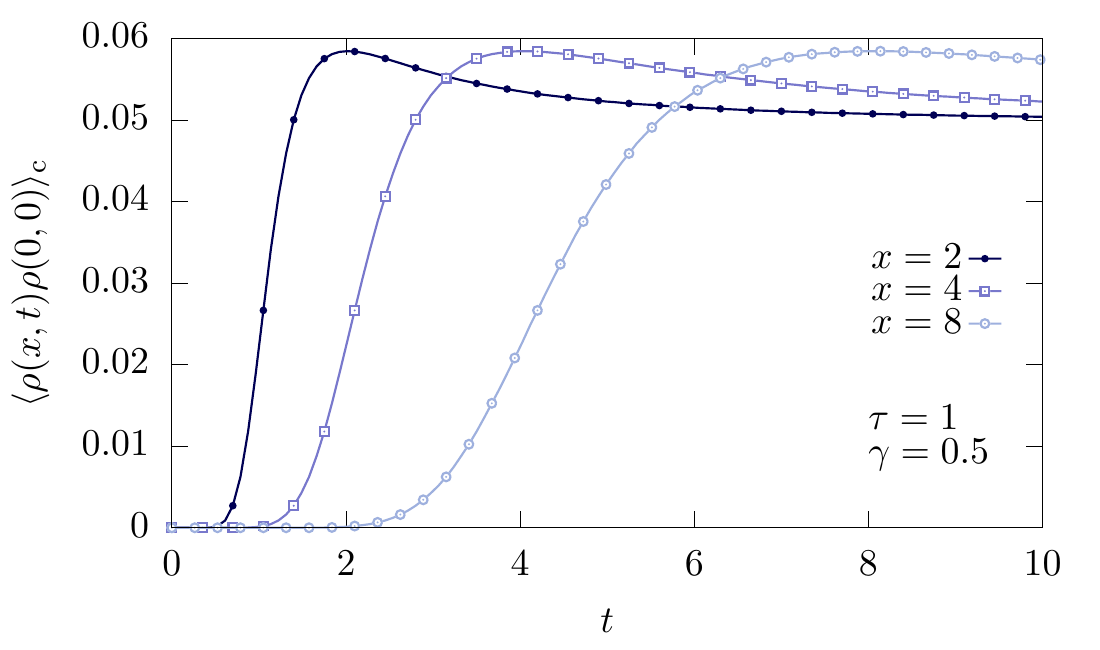}
	\caption{ Density-density connected correlators $\langle\rho(x,t)\rho(0,0)\rangle_\text{c}$ on thermal states, where $\rho(x,t)=\Psi^{\dagger}(x,t)\Psi(x,t)$. The correlators are computed using \eqref{eq:twopoint_Euler}. The plot shows the time dependence of the latter for different distances $x$. The emergence of a light-cone effect is clearly visible as $x$ is increased.}
	\label{fig:correlators}
\end{figure}

In this section we finally present an application of our result in the context of the recently introduced generalized hydrodynamics \cite{CaDY16,BCDF16}. The latter, is a novel approach to the non-equilibrium dynamics of integrable systems in non-homogeneous settings, which has recently attracted a lot of attention, due to its simplicity and many applications \cite{Kormos2018,Doyon17,DoyonSphon17,GHD3,GHD6,GHD7,GHD8,GHD10,F17,DS,ID117,DDKY17,DSY17,ID217,CDV17,BDWY17,mazza2018,BFPC18,Bas_Deluca_defhop,Bas_Deluca_defising,Alba18,Bas2018,PeGa17,MPLC18,CDDK17,BePC18,DeBD18}.

A prototypical situation which can be studied by the generalized hydrodynamics is given by the junction of two semi-infinite subsystems which are prepared in different macrostates, and suddenly joined together. At large time $t$ and distance $x$ from the junction, a quasi-stationary state emerges which can be locally described by a space- and time-dependent rapidity distribution function $\rho_{t,x}(\lambda)$, which acquires the semiclassical interpretation of a local density of particles. Quasi-local stationary states at different points in space and times are related by a continuity equation of the form \cite{CaDY16,BCDF16}
\be\label{eq:hydro_root}
\partial_t \rho_{t,x}(\lambda)+\partial_x\left(v^\text{eff}(\lambda)\rho_{t,x}(\lambda)\right)=0\,.
\ee
Here, $v^\text{eff}(\lambda)$ is the effective velocity which is defined as 
\be
v^\text{eff}(\lambda)=\frac{\partial_\lambda \epsilon^\text{dr}(\lambda)}{\partial_\lambda p^\text{dr}(\lambda)}\, ,
\ee
where $\epsilon(\lambda)$ and $p(\lambda)$ are the single particle energy and momentum respectively. Finally, the dressing operation on a given function $h(\lambda)$ is defined as
\be\label{eq:dr_def}
h^\text{dr}(\lambda)=h(\lambda)+\int_{-\infty}^\infty \frac{\dd \mu}{2\pi} \varphi_\text{LL}(\lambda-\mu)\vartheta(\mu)h^\text{dr}(\mu)\, .
\ee
Notice that $v^\text{eff}(\lambda)$ acquires a space/time dependence due to the dressing operation, where the filling $\vartheta$ must be of course computed with the local distribution functions $\rho_{t,x}(\lambda)$.

GHD has been firstly formulated to describe partitioning protocols \cite{CaDY16,BCDF16}, where the dynamics is ruled by an homogeneous Hamiltonian and the inhomogeneity is  restricted to the initial state, but subsequent developments even considered smooth inhomogeneities in the Hamiltonian itself \cite{GHD3}, adding suitable force terms to Eq. \eqref{eq:hydro_root}.
Of course, since within the GHD approximation local observables are computed as if the system was homogeneous, our result for $\langle (\Psi^\dagger)^K(\Psi)^K\rangle$ can be readily used to study inhomogenous profiles of the one-point functions \cite{BaPC18}. This is reported in subfigure $(b)$ of Fig.\ref{fig:quench_ness}.

Besides providing one-point functions in inhomogeneous setups, GHD also allows us to compute suitable connected correlation functions at the so called Eulerian scale \cite{Doyon17,DoyonSphon17}, namely large distance and time interval. In particular, in the Lieb-Liniger model the following formula was derived for the two point-function at the Eulerian scale  \cite{Doyon17}
\be
\langle \mathcal{O}(x,t)\mathcal{O}'(0,0)\rangle_\text{c} =\int_{-\infty}^\infty \dd \lambda \, \delta(x-v^\text{eff}(\lambda)t)\rho(\lambda)[1-\vartheta(\lambda)]V^{\mathcal{O}}(\lambda)V^{\mathcal{O}'}(\lambda)\,.
\label{eq:twopoint_Euler}
\ee
By mean of an explicit integration of the $\delta-$ function, we obtain a scaling function in terms of the ray $\zeta=x/t$
\be
t\langle \mathcal{O}(\zeta t,t)\mathcal{O}'(0,0)\rangle_\text{c} =\left[\frac{\rho(\lambda)[1-\vartheta(\lambda)]V^{\mathcal{O}}(\lambda)V^{\mathcal{O}'}(\lambda)}{\partial_\lambda v^\text{eff}v(\lambda)}\right]_{v^\text{eff}(\lambda)=\zeta} \,. \label{eq:twopoint_Euler_scal}
\ee

Above, the functions $V^\mathcal{O}(\lambda)$ are defined as follows.
Assume the GGE is described by an integral equation of the form \eqref{eq:betaGGE}. Then $V^\mathcal{O}$ is defined varying the expectation values of $\langle \mathcal{O}\rangle$ with respect to the GGE source $w(\lambda)$, namely
\be\label{eq:betader}
-\delta\langle \mathcal{O}\rangle\Big|_{\beta=1}=\int \dd \lambda\, \rho(\lambda)[1-\vartheta(\lambda)] V^\mathcal{O}(\lambda) (\delta w)^\text{dr}(\lambda)\, .
\ee
Note that here we assume to work in a regime of large distances and times, so that the validity of the hydrodynamic formalism is guaranteed. We refer to \cite{Doyon17} for a detailed discuss on the range of validity of \eqref{eq:twopoint_Euler}.

Being the variation $\delta w$ arbitrary, the above equation completely identifies $V^\mathcal{O}$. Two-point Eulerian correlation functions through GHD were initially formulated for the density of charges and currents in an homogeneous background \cite{DoyonSphon17}, but later their validity have been conjectured for arbitrary local operators and multi-point generalizations in inhomogeneous background \cite{Doyon17}. 
In all these cases, the GHD formulas need as an input $V^\mathcal{O}$.
Within classical integrable models, the GHD prediction for correlation functions has been numerically verified \cite{BDWY17}, where a local averaging on fluid cells has been understood to be necessary in order to ensure the validity of Eq. \eqref{eq:twopoint_Euler} (see Ref.~\cite{BDWY17} for more details).

In the Lieb-Liniger model, GHD correlators for one-point functions have already been investigated in Ref. \cite{Doyon17}, but the computation of $V^\mathcal{O}$ was based on the formulas of \cite{Pozs11}, where one-point functions are expressed in terms of multiple integrals, making the final result difficult to be evaluated in practice. Our result, instead, allows us to find efficient expressions for GHD correlators. We consign the necessary calculations to Appendix~\ref{app:hydro}, whereas here we simply report the final result. Denoting with $V^K(\lambda)$ the $V-$function associated with the operator $(\Psi^\dagger)^K(\Psi)^K$, we obtain the compact formula
\be
\sum_{n=0}^{\infty}X^n\frac{V^n(\lambda)}{(n!)^2(\kappa m)^n}=-\exp\left(\frac{1}{2\pi m\kappa}\sum_{m=1}^{+\infty} X^m \mathcal{B}_m\right)\left(\frac{1}{2\pi m\kappa}\sum_{j=1}^{+\infty} X^j j^{-1} \sum_{n=1}^{2j-1} d_n^{2j-1}(\lambda) b_n(\lambda)\right)\, .
\label{eq:V_gen}
\ee
where $d_n^j$ are solutions of the following set of integral equations

\be\label{eq:int_d1}
d^j_{2n}(\lambda)=\delta_{2n,j}+\int_{-\infty}^\infty \frac{\dd \mu}{2\pi}\vartheta(\mu)\{ \Gamma(\lambda-\mu)\vartheta(\mu)[d^j_{2n+3}(\mu)-2d_{2n+1}^j(\mu)]+\varphi_\text{LL}(\lambda-\mu)[d_{2n}^j(\mu)-d_{2n+2}^j(\mu)]\}
\ee

\be\label{eq:int_d2}
d^j_{2n+1}(\lambda)=\delta_{2n+1,j}+\int_{-\infty}^\infty \frac{\dd \mu}{2\pi}\vartheta(\mu)\{ \Gamma(\lambda-\mu)\vartheta(\mu)[-d_{2n+2}^j(\mu)]+\varphi_\text{LL}(\lambda-\mu)[d_{2n+1}^j(\mu)-d_{2n+3}^j(\mu)]\}
\ee

Eq. \eqref{eq:int_d1} and \eqref{eq:int_d2} can be solved recursively in analogy to Eqs.~\eqref{eq:int_1} and \eqref{eq:int_2}, but proceeding in the opposite direction: we fix $d_n^j(\lambda)=0$ for $n>j$, then keeping $j$ fixed Eq.\eqref{eq:int_d1} and \eqref{eq:int_d2} recursively determines $d^j_{l\le n}(\lambda)$ proceeding from larger to smaller values of $l$. 

We stress that the results presented in this section can be understood as a more efficient version of the ones derived in \cite{Doyon17}. The physical content is obviously the same: in particular, formula \eqref{eq:twopoint_Euler} is taken without modifications from \cite{Doyon17}, so that our contribution only amounts to a more efficient computation of the functions $V^{\mathcal{O}}$ for local operators. For completeness, we display in Fig.~\ref{fig:correlators} the two-point connected correlators for the most interesting case of the density operator $\rho(x,t)=\Psi^{\dagger}(x,t)\Psi(x,t)$, as computed from \eqref{eq:twopoint_Euler}. We see from the figure that a clear light-cone effect is emerging: the correlator $\langle\rho(x,t)\rho(0,0)\rangle_\text{c}$ is initially vanishing, and starts to deviate from zero only after a certain time interval which increases linearly as the distance $x$ increases. We verified that a similar qualitative behavior is obtained for higher local operators $\mathcal{O}=(\Psi^\dagger)^K(\Psi)^K$.

\section{Conclusions}
\label{sec:conclusions}

In this work we have derived analytic expressions for
the $n$-body local correlation functions for arbitrary macrostates in the Lieb-Liniger model, by exploiting the non-relativistic limit of the shG field theory. Most of our results were previously announced in \cite{BaPC18}; here a complete derivation was presented, together with a full survey of their physical applications, which include a computation of the full counting statistics for particle-number fluctuations. We have shown that our formulas are extremely convenient for explicit numerical computations, by presenting their evaluation for several physically interesting macrostates, including thermal states, GGEs and non-equilibrium steady states arising in transport problems. Furthermore, by building upon recent results within the framework of GHD, we provided efficient formulas for the computation of multi-point correlations at the Eulerian scale. Complementing previous studies in the literature, our results provide a full solution to the problem of computing one-point functions in the Lieb-Liniger model.

Our work shows once again the power of the non-relativistic limit first introduced in \cite{KoMT09} for the computation of local observables. Different important directions remain to be investigated. On the one hand, an interesting generalization of the LeClair-Mussardo series for non-local observables was derived in \cite{PoSz18}, and it is natural to wonder whether an appropriate non-relativistic limit could be performed to obtain analogous results also for the Lieb-Liniger gas. These would be extremely relevant in connection with cold-atom experiments. On the other hand, non-relativistic limits have been worked out also for other field-theories \cite{KoMP10,KoMT11,CaKL14,BaLM16,BaLM17}, and it is natural to wonder whether our results can be generalized. In particular, the most natural question pertains the sine-Gordon field theory, which is mapped onto the attractive one-dimensional Bose gas \cite{CaKL14,PiCE16}. 
Indeed, the techniques which eventually led to the Negro-Smirnov formula in the shG model were originally introduced in the sine-Gordon model \cite{JMS11,JMS11b}. However, in the sine-Gordon case only zero-temperature results were achieved so far \cite{JMS09}. We hope that our findings will motivate further studies in this direction.

\begin{acknowledgments} %
We acknowledge helpful discussions with Pasquale Calabrese, Bal\'azs Pozsgay and M\'arton Kormos.
\end{acknowledgments}%

\appendix

\section{Analytic test of the main result}
\label{app:checks}

In this appendix we show how to perturbatively test our main formula \eqref{eq:main_corr}, against previous results available in the literature. In particular, we compare our findings with those of \cite{Pozs11}, where $\mathcal{O}_n$ is explicitly worked out up to $n=4$. The results of \cite{Pozs11} could be summarized as follows. Define the auxiliary function $h^{(\ell)}(\lambda)$ by
\be
h^{(\ell)}(\lambda)=\lambda^{\ell}+\int_{-\infty}^{+\infty}\frac{\dd \mu}{2\pi}\,\varphi\left(\lambda-\mu\right)\vartheta(\mu)h^{\ell}(\mu)\,,
\ee
and
\be
\{n,m\}:=\int_{-\infty}^{+\infty}\frac{\dd}{2\pi}\vartheta(\mu)\mu^{n}h^{(m)}(\mu)\,.
\ee
Then, one has
\be
\mathcal{O}_2=\frac{2}{c} \Big( \{0,2\}-\{1,1\}\Big)\,,
\label{eq:g2_pozsgay}
\ee
\be
\begin{split}
	\mathcal{O}_3=&\frac{1}{c^2}  \Big(-4 \{1,3\}+3\{2,2\}+\{0,4\}\Big)
	+\Big(\{0,2\}-\{1,1\}\Big)
	+\frac{2}{c}\Big(\{0,1\}^2-\{0,0\}\{1,1\}\Big)\,,
\end{split}
\label{eq:g3_pozsgay}
\ee
and
\be
\begin{split}
	\mathcal{O}_4=\frac{2}{5c^3}\Big[&
	8 c^3 \Big(\{0, 1\}^2  -  \{0, 0\} \{1, 1\}\Big) +  
	32 c \Big(\{0, 1\} \{0, 3\}-  \{0, 0\} \{1, 3\}\Big) +\\
	& 24c\Big(\{0, 2\} \{1, 1\}- \{0, 1\}\{1,2\}\Big) 
	+ 30 c \Big(\{0, 0\} \{2, 2\}-\{0, 2\}^2\Big)+\\
	& 4 c^4 \Big(\{0, 2\} -\{1,1\}\Big) + 
	5c^2 \Big(\{0, 4\}  - 4 \{1, 3\}+3  \{2, 2\}\Big)+\\
	&  \{0, 6\}   - 6 \{1, 5\} + 15 \{2, 4\}  -  10 \{3, 3\} 
	\Big]\,.
	\label{eq:g4_pozsgay}	
\end{split}
\ee
From these expressions one can compute a perturbative expansion using the function $\vartheta(\lambda)$ as the small parameter, and compare every order with the analogous expansion obtained starting from \eqref{eq:main_corr}. More precisely, we have
\bea
\mathcal{O}_m=\sum_{n=0}^{\infty}\frac{1}{(2\pi)^n}\int\dd\mu_1\ldots\dd \mu_n \vartheta(\mu_1)\ldots \vartheta(\mu_n)f^{(m)}(\mu_1,\ldots,\mu_n)=\sum_{n=0}^{\infty}\mathcal{H}^{(m)}_n\,.
\eea
The terms $\mathcal{H}_n^{(m)}$ can be easily computed from \eqref{eq:g2_pozsgay}-\eqref{eq:g4_pozsgay}. For example, for $m=2$, we obtain
\bea
\mathcal{H}^{(2)}_0&=&\mathcal{H}_1=0\,,\label{eq:h0}\\
\mathcal{H}^{(2)}_{2}&=&\int\dd\mu_1\dd\mu_2\,h(\mu_1,\mu_2)=\frac{1}{2\pi^2\kappa}\int\dd\mu_1\dd\mu_2\left[\mu_1^2\varphi(\mu_1-\mu_2)-\mu_1\varphi(\mu_1-\mu_2)\mu_2\right]\,,\\
\mathcal{H}^{(2)}_{3}&=&\int\dd\mu_1\dd\mu_2\dd\mu_3\,h(\mu_1,\mu_2,\mu_3)=\frac{1}{4\pi^3\kappa}\int\dd\mu_1\dd\mu_2\dd\mu_3\left[\mu_1^2\varphi(\mu_1-\mu_2)\varphi(\mu_2-\mu_3)\right.\nonumber\\
&-& \left.\mu_1\varphi(\mu_1-\mu_2)\varphi(\mu_2-\mu_3)\mu_3\right]\,.\label{eq:h3}
\eea
An analogous expansion can be performed from \eqref{eq:main_corr}, as we now explicitly show for $m=2$. First we compute the following expansions, which can be obtained from \eqref{eq:int_1} and \eqref{eq:int_2}:
\bea
b_1(\lambda)&=&1+\frac{1}{2\pi}\int\dd\mu_1\vartheta(\mu_1)\varphi\left(\lambda-\mu_1\right)
+\frac{1}{(2\pi)^2}\int\dd\mu_1\dd\mu_2\vartheta(\mu_1)\vartheta(\mu_2)\varphi\left(\lambda-\mu_1\right)\varphi\left(\mu_1-\mu_2\right)+\ldots\,,\\
b_3(\lambda)&=&-\frac{1}{2\pi}\int\dd\mu_1\vartheta(\mu_1)\varphi\left(\lambda-\mu_1\right)
-\frac{2}{(2\pi)^2}\int\dd\mu_1\dd\mu_2\vartheta(\mu_1)\vartheta(\mu_2)\varphi\left(\lambda-\mu_1\right)\varphi\left(\mu_1-\mu_2\right)\nonumber\\
&+&\frac{2}{(2\pi)^2}\int\dd\mu_1\dd\mu_2\vartheta(\mu_1)\vartheta(\mu_2)\Gamma\left(\lambda-\mu_1\right)\Gamma\left(\mu_1-\mu_2\right)+\ldots \,,
\eea
so that
\bea
\mathcal{B}_1&=&\int\dd\mu_1\vartheta(\mu_1)+\frac{1}{2\pi}\int\dd\mu_1\dd\mu_2\vartheta(\mu_1)\vartheta(\mu_2)\varphi\left(\mu_1-\mu_2\right)\nonumber\\
&+&\frac{1}{(2\pi)^2}\int\dd\mu_1\dd\mu_2\dd\mu_3\vartheta(\mu_1)\vartheta(\mu_2)\vartheta(\mu_3)\varphi\left(\mu_1-\mu_2\right)\varphi\left(\mu_2-\mu_3\right)+\ldots\,,\\
\mathcal{B}_2&=&\frac{1}{2}\left[-\frac{1}{2\pi}\int\dd\mu_1\dd\mu_2\vartheta(\mu_1)\vartheta(\mu_2)\varphi\left(\mu_1-\mu_2\right)
-\frac{2}{(2\pi)^2}\int\dd\mu_1\dd\mu_2\dd\mu_3\vartheta(\mu_1)\vartheta(\mu_2)\vartheta(\mu_3)\varphi\left(\mu_1-\mu_2\right)\varphi\left(\mu_2-\mu_3\right)\nonumber\right.\\
&+&\left.\frac{2}{(2\pi)^2}\int\dd\mu_1\dd\mu_2\dd\mu_3\vartheta(\mu_1)\vartheta(\mu_2)\vartheta(\mu_3)\Gamma\left(\mu_1-\mu_2\right)\Gamma\left(\mu_2-\mu_3\right)\right]+\ldots\,.
\eea
Plugging these expressions into \eqref{eq:main_corr} we get
\be
\mathcal{O}_2=\mathcal{G}_0+\mathcal{G}_1+\mathcal{G}_2+\mathcal{G}_3+\ldots\,,
\ee
where
\bea
\mathcal{G}_0&=&\mathcal{G}_1=0\,,\label{eq:g1}\\
\mathcal{G}_2&=&\int\dd\mu_1\dd\mu_2\,g(\mu_1,\mu_2)=\frac{1}{2\pi^2}\int\dd\mu_1\dd\mu_2\vartheta(\mu_1)\vartheta(\mu_2)\left[1-\frac{\kappa}{2}\varphi(\mu_1-\mu_2)\right]\,,\\
\mathcal{G}_3&=&\int\dd\mu_1\dd\mu_2\dd\mu_3\,g(\mu_1,\mu_2,\mu_3)=\frac{1}{2\pi^2}\int\dd\mu_1\dd\mu_2\dd\mu_3\vartheta(\mu_1)\vartheta(\mu_2)\vartheta(\mu_3)\left[\frac{1}{2\pi}\varphi(\mu_1-\mu_2)+\frac{1}{2\pi}\varphi(\mu_2-\mu_3)\right.\nonumber\\
&+&\left.\kappa\pi(-\frac{2}{(2\pi)^2}\varphi\left(\mu_1-\mu_2\right)\varphi\left(\mu_2-\mu_3\right)+\frac{2}{(2\pi)^2}\Gamma\left(\mu_1-\mu_2\right)\Gamma\left(\mu_2-\mu_3\right))\right]\,.\label{eq:g3}
\eea
Comparing \eqref{eq:h0}-\eqref{eq:h3} with \eqref{eq:g1}-\eqref{eq:g3} we see that the two expansions are equal provided that the fully symmetrized functions obtained from $h(\mu_1,\ldots, \mu_r)$ and $g(\mu_1,\ldots, \mu_r)$ coincide, namely
\be
\sum_{\sigma \in S_r}h(\mu_{\sigma(1)},\ldots ,\mu_{\sigma( r)})=\sum_{\sigma \in S_r}g(\mu_{\sigma(1)},\ldots ,\mu_{\sigma(r)})\,,\qquad r=1,\ldots, 3\,,
\ee
where the sums are over all the permutations $\sigma$ of $r$ elements. One can see straightforwardly that this equation is verified. An analogous treatment can be done for $\mathcal{O}_n$ with $n\geq 3$, even though the calculations become increasingly cumbersome with $n$ and the order of the expansion.

\section{Proof that the coefficients $\mathcal{A}_j$ are vanishing}
\label{app:checkA}

In this appendix we show that the coefficients $\mathcal{A}_j$, defined in \eqref{eq:aj_coefficient}, are vanishing. This can be easily established by symmetry arguments in the case $\vartheta(\lambda)$ is a symmetric function of $\lambda$. This is true for thermal states, but not for quasi-stationary states arising in transport problems \cite{BCDF16}. A more sophisticated treatment is needed in the general case, which is sketched in the following. First, note that it is sufficient to show that
\be
\tilde{\mathcal{A}}_j:=\int_{-\infty}^\infty\dd\mu\, \vartheta(\mu)b_{2j}(\mu)=0\,.
\label{eq:to_prove}
\ee
Indeed, if this is true then multiplying both sides of \eqref{b_def} by $\vartheta(\lambda)$ and integrating in $\lambda$ we get immediately $\mathcal{A}_j=0$. Eq.~\eqref{eq:to_prove} can be established through a formal expansion of the functions $b_{j}(\lambda)$; we show this explicitly for $\mathcal{A}_1$, since an analogous treatment can be carried out for larger $j$. 

We start with the formal solution for the function $b_1(\lambda)$; from Eq.~\eqref{eq:int_1} with $n=0$ we have
\be
b_1(\lambda)=\sum_{n=0}^{\infty}\int\dd\mu_1\ldots\dd\mu_n\varphi(\lambda-\mu_1)\ldots\varphi(\mu_{n-1}-\mu_n)\vartheta(\mu_1)\ldots \vartheta(\mu_n)\,.
\ee
Next, plugging this into Eq.~\eqref{eq:int_2} for $n=1$ we obtain the formal solution
\bea
b_2(\lambda)&=&2\sum_{m=0}^{\infty}\sum_{n=0}^{\infty}\int\dd\nu_1\ldots \dd\nu_m\vartheta(\nu_1)\ldots\vartheta(\nu_m)\varphi(\lambda-\nu_1)\ldots \varphi(\nu_{m-1}-\nu_{m})\nonumber\\
&\times&\int\dd\mu_1\ldots \dd\mu_n\vartheta(\nu_1)\ldots \vartheta(\nu_n)\Gamma(\nu_m-\mu_0)\varphi(\mu_0-\mu_1)\ldots \varphi(\mu_{n-1}-\mu_{n})\,,
\eea
so that
\bea
\int \dd \lambda\,\vartheta(\lambda) b_2(\lambda)&=&\sum_{n=0}^{\infty}\sum_{m=0}^{\infty}\left(\prod_{j=1}^n\int \dd\mu_j \vartheta(\mu_j)\right)\left(\prod_{j=1}^m\int \dd\sigma_j \vartheta(\sigma_j)\right)\int \dd\nu\, \vartheta(\nu)\int \dd\lambda\, \vartheta(\lambda) \varphi(\lambda-\mu_1)\nonumber\\
&\times &\varphi(\mu_1-\mu_2)\ldots \varphi(\mu_{n-1}-\mu_n)\Gamma(\mu_n-\nu)\varphi(\nu-\sigma_1)\varphi(\sigma_1-\sigma_2)\ldots\varphi(\sigma_{m-1}-\sigma_m)\nonumber\\
&=&\sum_{n=0}^{\infty}\sum_{m=0}^{\infty}\mathcal{C}(n,m)\,.
\label{eq:intermediate_3}
\eea 
In order to show that the above expression is vanishing, we show
\be
\mathcal{C}(n,m)=-\mathcal{C}(m,n)\,.
\label{eq:relation_c}
\ee
This implies that $\mathcal{C}(n,n)=0$ and that all the other terms in the infinite sums \eqref{eq:intermediate_3} cancel each other out, as they are pairwise opposite. The proof of \eqref{eq:relation_c} amounts to a change of variables in the multiple integrals. We rename the variables as
\bea
\begin{cases}
	\sigma_m=\lambda^{\prime}\,,\\
	\sigma_{m-1}=\mu_1^{\prime}\,,\\
	\vdots \\
	\sigma_{1}=\mu_{m-1}^{\prime}\,,\\
	\nu=\mu_{m}^{\prime}\,,
\end{cases}
\eea
and also
\be
\begin{cases}
	\mu_n=\nu^{\prime}\,,\\
	\mu_{n-1}=\sigma_1^{\prime}\,,\\
	\vdots \\
	\mu_{1}=\sigma_{n-1}^{\prime}\,,\\
	\lambda=\sigma_{n}^{\prime}\,.
\end{cases}
\ee
Then
\bea
\mathcal{C}(n,m)&=&\sum_{n=0}^{\infty}\sum_{m=0}^{\infty}\left(\prod_{j=1}^n\int \dd\mu_j \vartheta(\mu_j)\right)\left(\prod_{j=1}^m\int \dd\sigma_j \vartheta(\sigma_j)\right)\int \dd\nu\, \vartheta(\nu)\int \dd\lambda\, \vartheta(\lambda) \varphi(\lambda-\mu_1)\nonumber\\
&\times &\varphi(\mu_1-\mu_2)\ldots \varphi(\mu_{n-1}-\mu_n)\Gamma(\mu_n-\nu)\varphi(\nu-\sigma_1)\varphi(\sigma_1-\sigma_2)\ldots \varphi(\sigma_{m-1}-\sigma_m)\nonumber\\
&=&\sum_{n=0}^{\infty}\sum_{m=0}^{\infty}\left(\prod_{j=1}^m\int \dd\mu^{\prime}_j \vartheta(\mu^{\prime}_j)\right)\left(\prod_{j=1}^n\int \dd\sigma^{\prime}_j \vartheta(\sigma^{\prime}_j)\right)\int \dd\nu^{\prime}\, \vartheta(\nu^{\prime})\int \dd\lambda^{\prime}\, \vartheta(\lambda^{\prime}) \varphi(\mu^{\prime}_1-\lambda^{\prime})\nonumber\\
&\times &\varphi(\mu^{\prime}_2-\mu^{\prime}_1)\ldots \varphi(\mu^{\prime}_{m}-\mu^{\prime}_{m-1})\Gamma(\nu^{\prime}-\mu^{\prime}_n)\varphi(\sigma^{\prime}_1-\nu^{\prime})\varphi(\sigma^{\prime}_2-\sigma^{\prime}_1)\ldots \varphi(\sigma^{\prime}_n-\sigma^{\prime}_{n-1})\,,
\eea
where in the r.h.s. we have rearranged the terms. Using now $\varphi(-\lambda)=\varphi(\lambda)$, $\Gamma(-\lambda)=-\Gamma(\lambda)$ and that the integration variables are dumb indices, we finally get \eqref{eq:relation_c}.

\section{Normal ordering of the moment-generating function of $\hat{N}_\Delta$}
\label{app:normal_order}

This appendix is devoted to a rigorous proof of Eq.~\eqref{eq:norm_ord_full}. In order to do this, we introduce a  lattice regularization of the continuous gas, with a lattice spacing $a$, similarly to what has been done in Ref. \cite{KoCC14,BaCS17}. Once the combinatorics has been carried out, we will take the limit $a\to 0$ and recover the continuous theory.

We start by introducing the discrete bosonic operators $\psi_j$ that satisfy bosonic commutation rules $[\psi_j,\psi^\dagger_{j'}]=\delta_{j,j'}$. The correspondence to extract the continuum limit is encoded in
\be
\psi_j\to a^{1/2}\Psi(aj)\label{discrete_map}\, .
\ee
In the following, we refer to Ref. \cite{KoCC14,BaCS17} for a detailed justification of such a limit, summarizing here only the main points. The validity of the mapping can be understood taking a many body test state $\ket{st^\text{d}}$ in the discrete model, in the assumption that the wave function has a well defined continuum limit. Thus, we introduce
\be
\ket{st^\text{d}}=a^{n/2}\sum_{\{j_i\}} \Phi(aj_1,...,a j_n)\psi^\dagger_{j_1}...\psi^\dagger_{j_n}\ket{0^\text{d}}\, ,
\ee
where the state $\ket{st^\text{d}}$ is the discrete regularization of $\ket{st^\text{c}}$, defined as
\be
\ket{st^\text{c}}=\int \dd^nx\, \Phi(x_1,...,x_n)\Psi^\dagger(x_1)...\Psi^\dagger(x_n)\ket{0^\text{c}}\, .
\ee
Of course, $\ket{0^\text{d}}$ and $\ket{0^\text{c}}$ are, respectively, the vacuum in the discrete and continuous model.
Notice that a trivial substitution in the wave function $\psi_j\to a^{1/2}\Psi(aj)$ matches $\ket{st^\text{d}}\to \ket{st^\text{c}}$, but such a replacement is not rigorous and the mapping should be understood in a weak sense, at the level of expectation values.
For example, the norm of the state
\be
\langle st^\text{d}|st^\text{d}\rangle=n!a^n\sum_{\{j_i\}}|\Phi(aj_1,...,aj_n)|^2\to n!\int \dd^nx\, |\Phi(x_1,...,x_n)|^2=\langle st^\text{c}|st^\text{c}\rangle\, ,
\ee
where in the limit $a\to 0$ we replaced summations with integrals.
By mean of similar calculations, we can consider simple observables in the form $\sum_j t(aj) \psi^\dagger_j\psi_j$ with $t(x)$ a smooth function. It holds:
\be
\langle st^\text{d}|\sum_j t(aj) \psi^\dagger_j\psi_j|st^\text{d}\rangle\to \langle st^\text{c}|\int \dd x\, t(x) \Psi^\dagger(x)\Psi(x)|st^\text{c}\rangle\, ,
\ee
that justifies the map \eqref{discrete_map} at the level of observables
\be
\sum_j t(aj) \psi^\dagger_j\psi_j\to \sum_j t(aj) a \Psi^\dagger(aj)\Psi(aj)\simeq \int \dd x\,  t(x)\Psi^\dagger(x)\Psi(x)\, .
\ee

This exercise can be carried out for other operators too, leading to the same conclusions. Thus, rather than considering $\exp\left(\int_0^\Delta \dd x\, \Psi^\dagger(x)\Psi(x)\right)$, we study its discrete version
\be
e^{i\gamma\sum_{j=0}^{\Delta/a}\psi_j^\dagger\psi_j}=\prod_{j=0}^{\Delta/a}e^{i\gamma\psi_j^\dagger\psi_j}\, .
\ee
Our goal is now to put the above in normal order.
Since at different sites the bosonic operators commute, we can analyze each site separately and consider $e^{i\gamma\psi^\dagger_j\psi_j}$ for a given $j$. In the forthcoming calculations, since we are reasoning at fixed lattice site, we simply drop the index $j$. What we are aiming for is an expression of this form
\be\label{eq:C_exp}
e^{i\gamma\psi^\dagger\psi}=\sum_{l=0}^\infty C_l :(\psi^\dagger\psi)^l:\, .
\ee
In order to determine $C_l$, we compare the two sides on test states $\ket{n}=(\psi^\dagger)^n\ket{0}$
\be
e^{i\gamma\psi^\dagger\psi}\ket{n}=e^{i\gamma n}\ket{n}\, \hspace{2pc}
\sum_{l=0}^\infty C_l :(\psi^\dagger\psi)^l:\ket{n}=\sum_{l\le n}^\infty C_l \frac{n!}{(n-l)!}\ket{n}\, .
\ee
The $C_l$ coefficients are therefore the solution of
\be
e^{i\gamma n}=\sum_{l\le n}C_l\frac{n!}{(n-l)!}\,.
\ee
In order to solve this equation we introduce an auxiliary parameter $q$, multiply for $q^n/n!$ both sides and then sum over $n$
\be
\sum_{n=0}^\infty \frac{q^n}{n!}e^{i\gamma n}=\sum_{n=0}^\infty\sum_{l\le n}C_l\frac{1}{(n-l)!}\frac{q^n}{n!}\, .
\ee
The summation over $n$ on both sides is immediately performed and we get
\be
e^{qe^{i\gamma}}=e^{q}\sum_{l=0}^\infty C_l q^l\, \hspace{2pc}\Longrightarrow\hspace{2pc}
C_l=\frac{1}{l!}\left(e^{i\gamma}-1\right)^l\, .
\ee
Inserting this in \eqref{eq:C_exp} we obtain
\be
e^{i\gamma\psi^\dagger\psi}=\sum_{l=0}^\infty \frac{\left(e^{i\gamma}-1\right)^l}{l!}:(\psi^\dagger\psi)^l:=:e^{\left(e^{i\gamma}-1\right)\psi^\dagger\psi}:\, .
\ee
Extending this identity to several sites we finally arrive at
\be
e^{i\gamma\sum_{j=0}^{\Delta/a}\psi_j^\dagger\psi_j}=:\exp\left[\left(e^{i\gamma}-1\right)\sum_{j=0}^{\Delta/a}\psi_j^\dagger\psi_j\right]:\, ,
\ee
whose continuum limit is the desired identity \eqref{eq:norm_ord_full}.

\section{Hydrodynamic correlators}
\label{app:hydro}

In this Appendix we derive the kernels needed in the computation of the Eulerian correlators, namely the $V^K(\lambda)$ functions Eq. \eqref{eq:V_gen}, \eqref{eq:int_d1} and \eqref{eq:int_d2}.
Aiming for a direct application of the definition Eq. \eqref{eq:betader}, we vary in $\delta w$ both sides of the generating function \eqref{eq:generating_function}, obtaining
\be\label{der_gen}
\sum_{n=0}^{\infty}X^n\frac{\delta\langle (\psi^\dagger)^n(\psi)^n\rangle}{(n!)^2(\kappa m)^n}=\exp\left(\frac{1}{2\pi m\kappa}\sum_{m=1}^{+\infty} X^m \mathcal{B}_m\right)\left(\frac{1}{2\pi m\kappa}\sum_{j=1}^{+\infty} X^j \delta\mathcal{B}_j\right)\, .
\ee
From this expression we eventually extract the generating function for the $V^K(\lambda)$ kernels.
From the direct definition of $\mathcal{B}_j$ \eqref{eq:B_def}, we find 
\be\label{der_B}
\delta\mathcal{B}_j=j^{-1}\int_{-\infty}^\infty \dd \lambda\, [\delta\vartheta(\lambda)]b_{2j-1}(\lambda)+j^{-1}\int_{-\infty}^\infty \dd \lambda\, \vartheta(\lambda)[\delta b_{2j-1}(\lambda)]\,.
\ee
The simpler term is $\delta \vartheta$, which by means of Eq.~\eqref{eq:filling} can be rewritten as
\be\label{der_vartheta}
\delta \vartheta(\lambda)=-\vartheta^2(\lambda)e^{\varepsilon(\lambda)}\delta \varepsilon(\lambda)=\vartheta(\lambda)[\vartheta(\lambda)-1]\delta \varepsilon(\lambda)\, .
\ee
Next, varying both sides of Eq.~\eqref{eq:betaGGE} and comparing with the definition of the dressing Eq.~\eqref{eq:dr_def}, we readily discover $\delta \varepsilon(\lambda)=(\delta w)^\text{dr}(\lambda)$ which implies
\be\label{eq:dervarth}
\delta\vartheta(\lambda)=\vartheta(\lambda)[\vartheta(\lambda)-1](\delta w)^\text{dr}(\lambda)\, .
\ee
In order to study $\delta b_j(\lambda)$, it is useful to introduce an operatorial notation for the integral equations \eqref{eq:int_1} and \eqref{eq:int_2}. First, note that they can be written in compact notation as
\be\label{eq:int}
b_i(\lambda)=\delta_{i,1}+\int_{-\infty}^\infty \frac{\dd \mu}{2\pi} \, \vartheta(\mu) \left[\Gamma(\lambda-\mu)\sum_{j=1}^\infty U_{i,j} b_j(\mu)+\varphi(\lambda-\mu)\sum_{j=1}^\infty W_{i,j} b_j(\mu)\right]\,,
\ee
where the matrices $W_{i,j}$ and $U_{i,j}$ are defined as
\be
U_{i,j}=\begin{cases} 2\delta_{i-1,j}-\delta_{i-3,j}\hspace{1pc}&i\text{ even}\,,\\
	\delta_{i-1,j} \hspace{1pc}&i\text{ odd}\,,\end{cases} \hspace{3pc} W_{i,j}= \delta_{i,j}-\delta_{i-2,j}\, .
\label{eq:UW_def}
\ee
Next, we organize the functions $b_j(\lambda)$ in a single vector $[\vec{b}(\lambda)]_j=b_j(\lambda)$ and rewrite Eq. \eqref{eq:int} as

\be\label{eq:int_op}
\vec{b}(\lambda)=\vec{s}+\int_{-\infty}^\infty \frac{\dd\mu}{2\pi} \Gamma(\lambda-\mu)\vartheta(\mu)U\vec{b}(\mu)+\varphi_\text{LL}(\lambda-\mu)\vartheta(\mu)W\vec{b}(\mu)\, ,
\ee
where the source term is of course $[\vec{s}]_j=\delta_{j,1}$.
We can even push further the operatorial notation and look at the integrations as matrix products. In this respect, we introduce operators 
\be
\hat{\Gamma}_{\lambda,\mu}=\Gamma(\lambda-\mu)\, , \hspace{2pc}\hat{\varphi}_{\lambda,\mu}=\varphi_\text{LL}(\lambda-\mu)\, , \hspace{2pc} \hat{\vartheta}_{\lambda,\mu}=\delta(\lambda-\mu)\vartheta(\mu)\,,
\ee
and
\be
\hat{U}_{\lambda,\mu}=\delta(\lambda-\mu)U\, , \hspace{5pc} \hat{W}_{\lambda,\mu}=\delta(\lambda-\mu)W\,,
\ee
while we can think of $\vec{s}$ and $\vec{b}$ as vectors in this space. Matrix products are performed through integrations
\be
[O O']_{\lambda,\mu}=\int_{-\infty}^\infty \dd \xi\, O_{\lambda,\xi}O'_{\xi,\mu}\, .
\ee

In this notation, we rewrite Eq. \eqref{eq:int_op} as
\be
\vec{b}=\vec{s}+\frac{1}{2\pi} \big(\hat{\Gamma}\hat{\vartheta}\hat{U}+\hat{\varphi}\hat{\vartheta}\hat{W}\big)\vec{b}\, .
\ee
The formal solution is
\be\label{eq:form_sol}
\vec{b}=\Big[1-\frac{1}{2\pi} \big(\hat{\Gamma}\hat{\vartheta}\hat{U}+\hat{\varphi}\hat{\vartheta}\hat{W}\big)\Big]^{-1}\vec{s}\,,
\ee
which now we vary with respect to $w(\lambda)$:
\be
\delta\vec{b}=\frac{1}{2\pi}\Big[1-\frac{1}{2\pi} \big(\hat{\Gamma}\hat{\vartheta}\hat{U}+\hat{\varphi}\hat{\vartheta}\hat{W}\big)\Big]^{-1}\left(\hat{\Gamma}\hat{U}+\hat{\varphi}\hat{W}\right)\delta \hat{\vartheta}\Big[1-\frac{1}{2\pi} \big(\hat{\Gamma}\hat{\vartheta}\hat{U}+\hat{\varphi}\hat{\vartheta}\hat{W}\big)\Big]^{-1}\vec{s}\, .
\ee

Using Eq. \eqref{eq:dervarth} together with Eq. \eqref{eq:form_sol}, the above can be written as
\be
\delta\vec{b}=\frac{1}{2\pi}\Big[1-\frac{1}{2\pi} \big(\hat{\Gamma}\hat{\vartheta}\hat{U}+\hat{\varphi}\hat{\vartheta}\hat{W}\big)\Big]^{-1}\left(\hat{\Gamma}\hat{U}+\hat{\varphi}\hat{W}\right)\hat{\vartheta}[\hat{\vartheta}-1](\hat{\delta w})^\text{dr}\vec{b}\, .
\ee
Equivalently, we can recast the above as
\be
\delta\vec{b}=-[\hat{\vartheta}-1]\hat{w}^\text{dr}\vec{b}+\Big[1-\frac{1}{2\pi} \big(\hat{\Gamma}\hat{\vartheta}\hat{U}+\hat{\varphi}\hat{\vartheta}\hat{W}\big)\Big]^{-1}[\hat{\vartheta}-1](\hat{\delta w})^\text{dr}\vec{b}\,,
\ee

Now, in $\delta\mathcal{B}_j$ we actually need $\int_{-\infty}^\infty \dd\lambda \vartheta(\lambda)\delta b_{2j-1}(\lambda)$, thus we contract with $\vartheta$ the above and rewrite it as
\be
\hat{\vartheta}\delta\vec{b}=-\hat{\vartheta}[\hat{\vartheta}-1]\hat{w}^\text{dr}\vec{b}+\Big[1-\frac{1}{2\pi} \big(\hat{\vartheta}\hat{\Gamma}\hat{U}+\hat{\vartheta}\hat{\varphi}\hat{W}\big)\Big]^{-1}\hat{\vartheta}[\hat{\vartheta}-1](\hat{\delta w})^\text{dr}\vec{b}\,.
\ee
We can now finally compute $\int_{-\infty}^\infty \dd \lambda \,\vartheta(\lambda)\delta\vec{b}(\lambda)$. Making the integrations explicit we have
\bea
\label{eq:D15}
\int_{-\infty}^\infty \dd \lambda \vartheta(\lambda)\delta\vec{b}(\lambda)&=&-\int_{-\infty}^\infty \dd \lambda\vartheta(\lambda)[\vartheta(\lambda)-1](\delta w)^\text{dr}(\lambda)\vec{b}(\lambda)\nonumber\\
&+&\int_{-\infty}^\infty \dd\xi\dd\lambda\, \Big[1-\frac{1}{2\pi} \big(\hat{\vartheta}\hat{\Gamma}\hat{U}+\hat{\vartheta}\hat{\varphi}\hat{W}\big)\Big]^{-1}_{\xi,\lambda}\vartheta(\lambda)[\vartheta(\lambda)-1](\delta w)^\text{dr}(\lambda)\vec{b}(\lambda)\,.
\eea
Let us now define the operator $D_{\xi,\lambda}$ as
\be
D=\Big[1-\frac{1}{2\pi} \big(\hat{\vartheta}\hat{\Gamma}\hat{U}+\hat{\vartheta}\hat{\varphi}\hat{W}\big)\Big]^{-1}\,,
\ee
which of course satisfies the equation
\be\label{eq_intD}
D=1+\frac{1}{2\pi}D\big(\hat{\vartheta}\hat{\Gamma}\hat{U}+\hat{\vartheta}\hat{\varphi}\hat{W}\big)\,.
\ee
As it is clear from Eq. \eqref{eq:D15}, we ultimately need $\int_{-\infty}^\infty \dd \xi\, D_{\xi,\lambda}$. Thus, we define
\be
\int_{-\infty}^\infty \dd \xi\, [D_ {\xi,\lambda}]_{j,n}=d_n^j(\lambda)\,.
\ee
From this definition and Eq. \eqref{eq_intD} we readily get a set of integral equations for $d_n^j(\lambda)$
\be
d_n^j(\lambda)=\delta_{n,j}+\int_{-\infty}^\infty \frac{\dd \mu}{2\pi} \sum_{l=1}^\infty d_l^j(\mu)\vartheta(\mu)\Big[\Gamma(\mu-\lambda)U_{l,n}+\varphi_\text{LL}(\mu-\lambda)W_{l,n}\Big]\,.
\ee

Exploiting the symmetries of the kernels and making explicit the matrix elements, this equation is seen to be identical to Eq. \eqref{eq:int_d1} and \eqref{eq:int_d2}. Making use of the functions $d_n^j(\lambda)$ defined in Eq. \eqref{eq:D15} we are finally led to
\be
\int_{-\infty}^\infty \dd \lambda \,\vartheta(\lambda)\delta b_j(\lambda)=\int_{-\infty}^\infty \dd \lambda\, \Big[-\vartheta(\lambda)[\vartheta(\lambda)-1](\delta w)^\text{dr}(\lambda)b_j(\lambda)+\sum_{n=1}^\infty d_n^j(\lambda)\vartheta(\lambda)[\vartheta(\lambda)-1](\delta w)^\text{dr}(\lambda)b_n(\lambda)\Big]\,.
\label{eq:temp}
\ee
Notice that, as we commented below Eq. \eqref{eq:int_d1} and \eqref{eq:int_d2}, we have $d_{n>j}^j(\lambda)=0$ thus the above series is truncated to a simple sum. Finally, plugging \eqref{eq:temp} and \eqref{eq:dervarth} into \eqref{der_B} we get
\be
\delta\mathcal{B}_j=j^{-1}\int_{-\infty}^\infty \dd \lambda\, \sum_{n=1}^{2j-1} d_n^{2j-1}(\lambda) b_n(\lambda)\vartheta(\lambda)[\vartheta(\lambda)-1](\delta w)^\text{dr}(\lambda)\,.
\ee
Inserting this result in Eq. \eqref{der_gen} and comparing with the definition Eq.\eqref{eq:betader}, we immediately obtain the desired result~\eqref{eq:V_gen}.

\end{document}